\documentclass[aps,prb,groupaddress,floatfix, twocolumn]{revtex4}

\usepackage{graphicx,graphics}
\usepackage{dcolumn}
\usepackage{amsmath,amssymb,amsfonts}
\usepackage{latexsym,verbatim}
\usepackage{bm}
\usepackage{color}
\usepackage{ulem}
\def\kv{{\bm k}}
\def\qv{{\bm q}}
\def\pv{{\bm p}}
\def\sigmav{{\bm \sigma}}
\newcommand{\h}[1]{{\hat {#1}}}
\newcommand{\hdg}[1]{{\hat {#1}^\dagger}}
\newcommand{\bra}[1]{\left\langle{#1}\right|}
\newcommand{\ket}[1]{\left|{#1}\right\rangle}
\def\be{\begin{equation}}
\def\ee{\end{equation}}
\def\ber{\begin{eqnarray}}
\def\eer{\end{eqnarray}}
\def\nn{\nonumber}

\begin{document}
\title{Drude weight, plasmon dispersion, and a.c. conductivity in doped graphene sheets}
\author{Saeed H. Abedinpour}
\thanks{Present address: Department of Physics, Institute for Advanced Studies in Basic Sciences (IASBS), Zanjan 45137, Iran}
\affiliation{Department of Physics and Astronomy, University of Missouri, Columbia, Missouri 65211, USA}
\author{G. Vignale}
\affiliation{Department of Physics and Astronomy, University of Missouri, Columbia, Missouri 65211, USA}
\author{A. Principi}
\affiliation{NEST, Istituto Nanoscienze-CNR and Scuola Normale Superiore, I-56126 Pisa, Italy}	
\author{Marco Polini}
\email{m.polini@sns.it}\homepage{http://qti.sns.it}
\affiliation{NEST, Istituto Nanoscienze-CNR and Scuola Normale Superiore, I-56126 Pisa, Italy}
\author{Wang-Kong Tse}
\affiliation{Department of Physics, University of Texas at Austin, Austin, Texas 78712, USA}
\author{A.H. MacDonald}
\affiliation{Department of Physics, University of Texas at Austin, Austin, Texas 78712, USA}
\begin{abstract}
We demonstrate that the plasmon frequency and Drude weight of the electron liquid in a doped graphene sheet are strongly renormalized by electron-electron interactions even in the long-wavelength limit. This effect is not captured by the Random Phase Approximation (RPA), commonly used to describe electron fluids and is due to coupling between the center of mass motion and the pseudospin degree of freedom of the graphene's massless Dirac fermions.  Making use of diagrammatic perturbation theory to first order in the electron-electron interaction, we show that this coupling {\it enhances} both the plasmon frequency and the Drude weight relative to the RPA value. We also show that interactions are responsible for a significant enhancement of the optical conductivity at frequencies just above the absorption threshold.   Our predictions can be checked by far-infrared spectroscopy or inelastic light scattering.
\end{abstract}

\maketitle

\section{Introduction}
\label{sect:intro}

The first theory of classical collective electron density oscillations in ionized gases 
by Tonks and Langmuir~\cite{tonks_langmuir_pr_1929} in the 1920's helped initiate the field of 
plasma physics.  The theory of collective electron density oscillations in metals, quantum in this case because of higher electron densities, 
was developed by Bohm and Pines~\cite{pines_bohm_pr_1952,Pines_and_Nozieres} in the 1950's and stands as a similarly pioneering contribution to many-electron physics.  Bohm and Pines coined the term {\it plasmon} to describe quantized density oscillations.    
Today {\it plasmonics} is a very active subfield of optoelectronics~\cite{Ebbesen_PT_2008,Maier07},  
whose aim is to exploit plasmon properties in order to compress infrared electromagnetic waves to the nanometer
scale of modern electronic devices.  This wide importance of plasmons across different fields of 
basic and applied physics follows from the ubiquity of 
charged particles and from the strength of their long-range Coulomb interactions.  

The physical origin of plasmons is very simple. When electrons in free space 
move to screen a charge inhomogeneity, they tend to overshoot the mark. They are then pulled back toward the charge disturbance and overshoot again, setting up a weakly damped oscillation.
The restoring force responsible for the oscillation is the average self-consistent field created by all the electrons. Because of the long-range nature of the Coulomb interaction, the frequency of oscillations $\omega_{\rm pl}(q)$ tends to be high and is given in the long wavelength limit by $\omega^2_{\rm pl}(q\to 0)= n q^2 V_q/m$ where $n$ is the electron density, $m$ is the bare electron mass in vacuum, and $V_q$ is the Fourier transform of the Coulomb interaction.  This simple explicit plasmon energy expression 
is exact because long-wavelength plasmons involve rigid motion of the entire plasma, which does not involve  
the complex exchange and correlation effects that dress~\cite{Giuliani_and_Vignale} the motion of an individual electron. 
The exact plasmon frequency expression is correctly captured by the  RPA~\cite{Pines_and_Nozieres,pines_bohm_pr_1952,Giuliani_and_Vignale}, but also by rigorous arguments~\cite{morchio_strocchi_ap_1986} in which the selection of a particular center-of-mass position breaks the system's Galilean invariance and 
plasmon excitations play the role of Goldstone bosons.  In two-dimensional (2D) systems $V_q =2\pi e^2/q$ so that $\omega_{\rm pl}(q\to 0) = \sqrt{2\pi n e^2q/m}$, where $e$ is the magnitude of the electron charge.

Electrons in a solid, unlike electrons in a plasma or electrons with a {\it jellium model}~\cite{Giuliani_and_Vignale} background,
experience a periodic external potential created by the ions which breaks translational invariance and hence also Galilean invariance.
Solid state effects can lead in general to a renormalization of the plasmon frequency, or even to the absence of sharp plasmonic excitations. 
In semiconductors and semimetals, however, electron waves can be described at super-atomic length scales using $\kv \cdot \pv$ theory~\cite{Cardona}, which is based on an expansion of the crystal's Bloch Hamiltonian around band extrema.  In the simplest case, for example for the conduction band of common cubic semiconductors, this leads us back to a Galilean-invariant parabolic band continuum model with isolated electron energy $E_{\rm c}(\pv) = \pv^2/(2m_{\rm b})$.  The crystal background for electron waves appears only via the replacement of the bare electron mass by an effective band mass $m_{\rm b}$.  In this type of $\kv \cdot \pv$ Galilean-invariant interacting electron model, valid for many semiconductor and semiconductor heterojunction systems, the plasmon dispersion is accurately given by the random phase approximation (RPA)~\cite{Pines_and_Nozieres,pines_bohm_pr_1952,Giuliani_and_Vignale} and 
it is given by the classical formula quoted above with the replacements $m \to m_{\rm b}$ and $e^2 \to e^2/\epsilon$, $\epsilon$ being the high-frequency dielectric constant of the semiconductor material.  The absence of electron-electron interaction corrections to plasmon frequencies at very long wavelengths  in these systems, has been demonstrated experimentally by means of inelastic light scattering~\cite{vittorio_ils,Hirjibehedin_prb_2007}.

The situation turns out to be quite different in graphene -- 
a monolayer of carbon atoms tightly packed in a 2D honeycomb lattice~\cite{geim_novoselov_nat_mat_2007,katsnelson_ssc_2007,allan_pt_2007,castro_neto_rmp_2009}.
When $\kv \cdot \pv$ theory is applied to graphene it leads to a new type of electron fluid model, one with separate Dirac-Weyl Hamiltonians for electron waves centered in momentum space on one of two honeycomb lattice Brillouin-zone corners $K$ and $K'$:
\be
{\hat {\cal H}}_{\rm D} = \hbar v \sum_{\kv, \alpha, \beta} {\hat \psi}^\dagger_{\kv, \alpha} \left(\sigmav_{\alpha\beta} 
\cdot \kv\right) {\hat \psi}_{\kv, \beta}~.
\ee 
Here $v$ is the bare electron velocity, $\kv$ is the $\kv \cdot \pv$ momentum, $\alpha,\beta$ are sublattice pseudospin labels,
and $\sigmav_{\alpha\beta} = (\sigma^x_{\alpha\beta}, \sigma^y_{\alpha\beta})$ is a vector of Pauli matrices which act on the sublattice pseudospin degree-of-freedom. It follows that the energy eigenstates for a given $\pv$ have pseudospins oriented either parallel (upper band) or antiparallel (lower band) to $\pv$.  Physically, the orientation of the pseudospin determines the relative amplitude and the relative phase of electron waves on the two distinct graphene sublattices. 

Electron-electron interactions in graphene 
are described by the usual non-relativistic Coulomb Hamiltonian~\cite{kotov_arXiv_2010}
\be\label{eq:Coulomb}
{\hat {\cal H}}_{\rm C} = \frac{1}{2S} \sum_{\qv \neq {\bm 0}} \; V_q \; {\hat \rho}_\qv \, {\hat \rho}_{- \qv}~,
\ee
where $S$ is the sample area, $V_q=2\pi e^2/(\epsilon q)$ is the 2D Fourier transform of the Coulomb interaction ($\epsilon$ being an effective average dielectric constant),  and ${\hat \rho}_\qv = \sum_{\kv, \alpha} {\hat \psi}^\dagger_{\kv - \qv, \alpha}{\hat \psi}_{\kv, \alpha}$ is the usual density operator. Electron carriers with density $n$ can be induced in graphene by purely electrostatic means, creating a circular 2D Fermi surface in the conduction band with a Fermi radius $k_{\rm F}$, which is proportional to $\sqrt{n}$$^{\,}$\cite{electron_doping}. 
The model described by ${\hat {\cal H}} = {\hat {\cal H}}_{\rm D} + {\hat {\cal H}}_{\rm C}$ requires an ultraviolet wavevector cutoff,
 $k_{\rm max}$, which should be assigned a value corresponding to the
wavevector range over which ${\hat {\cal H}}_{\rm D}$ describes graphene's $\pi$ bands. This corresponds to taking 
$k_{\rm max} \sim 1/a_0$ where $a_0 \sim 1.42$~\AA~is the carbon-carbon distance. This model is useful when $k_{\rm max}$ is much larger than $k_{\rm F}$. 

\begin{figure}
\tabcolsep=0cm
\begin{tabular}{cc}
\includegraphics[width=0.50\linewidth]{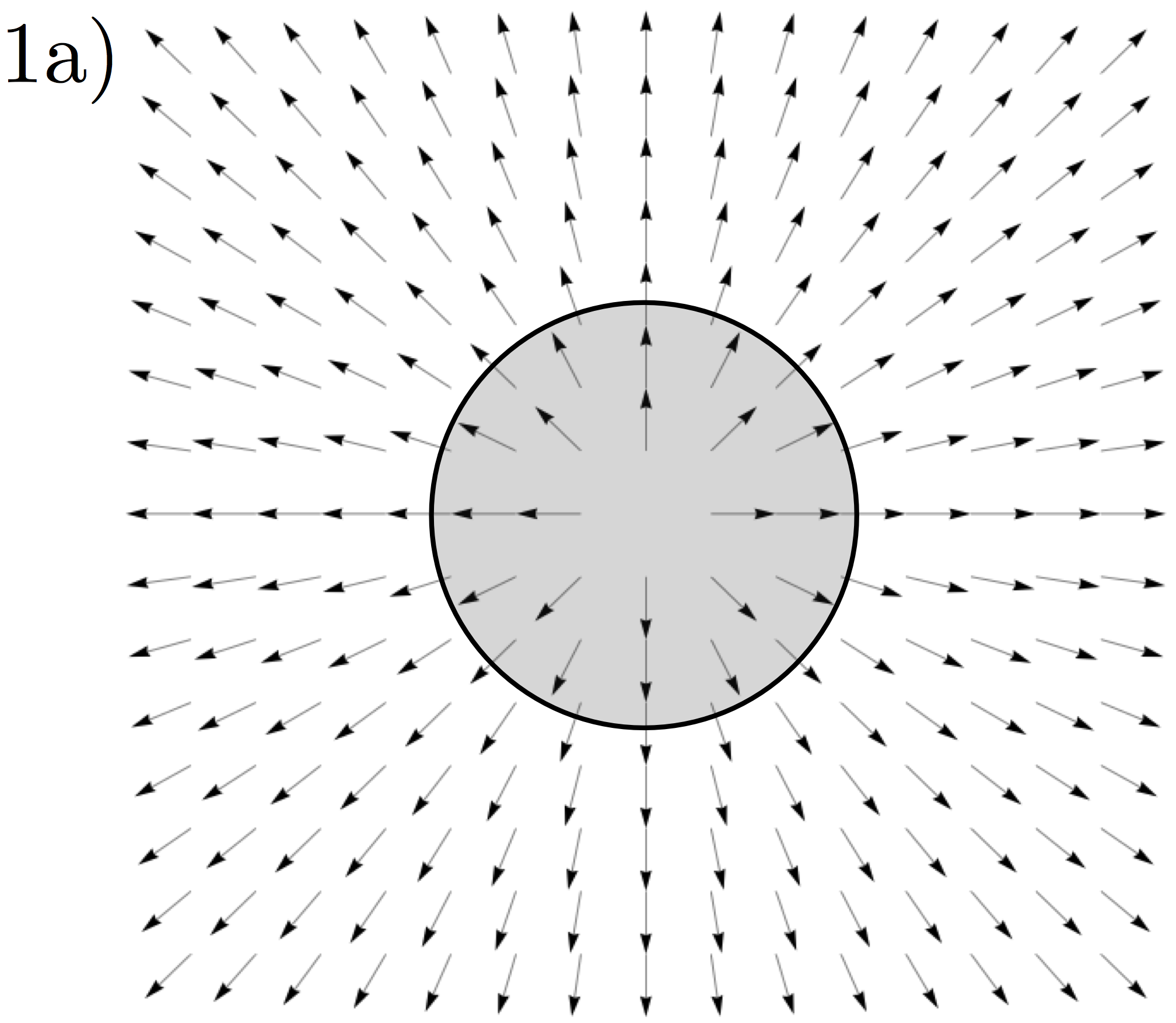} &
\includegraphics[width=0.50\linewidth]{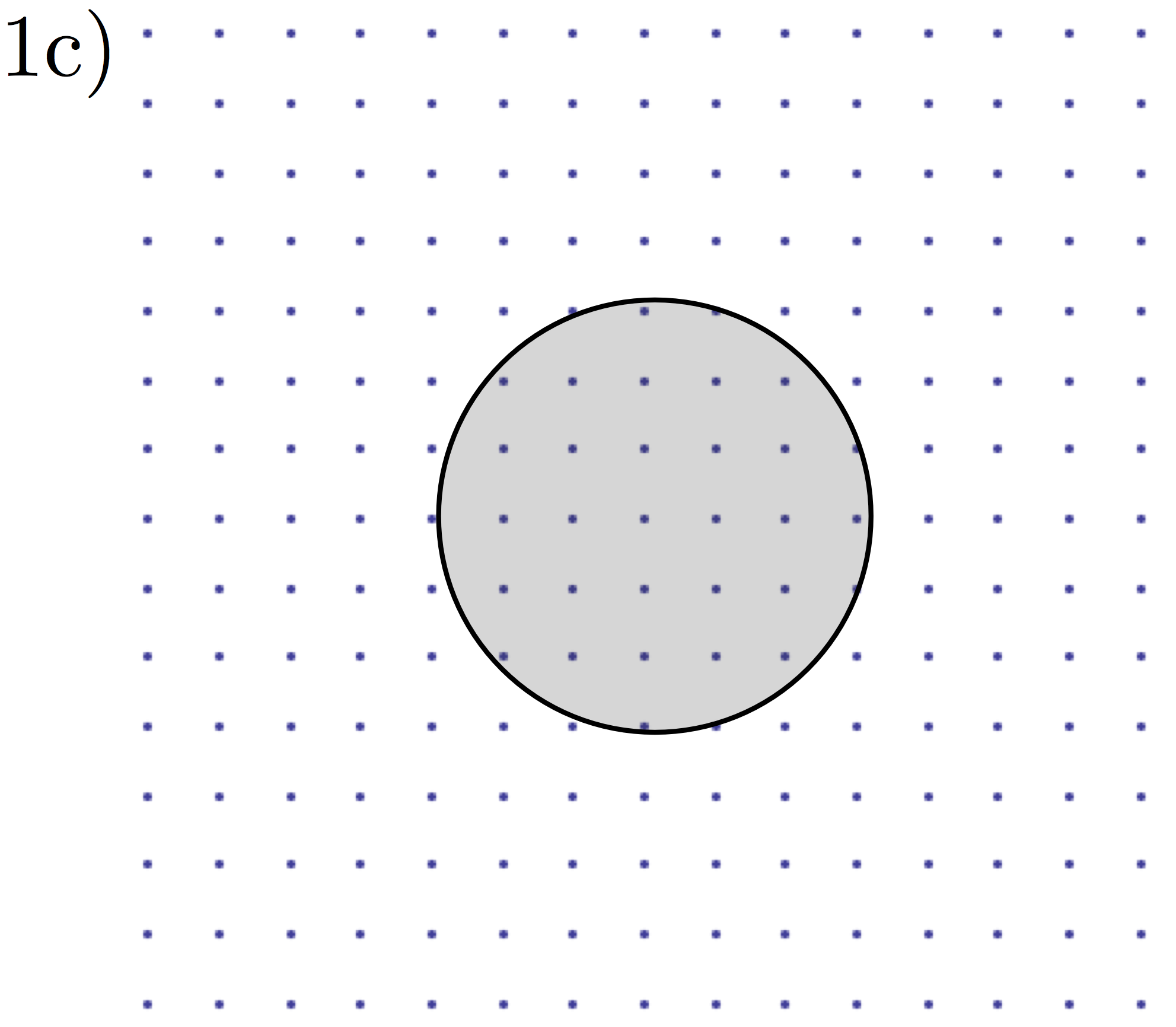}\\
\includegraphics[width=0.50\linewidth]{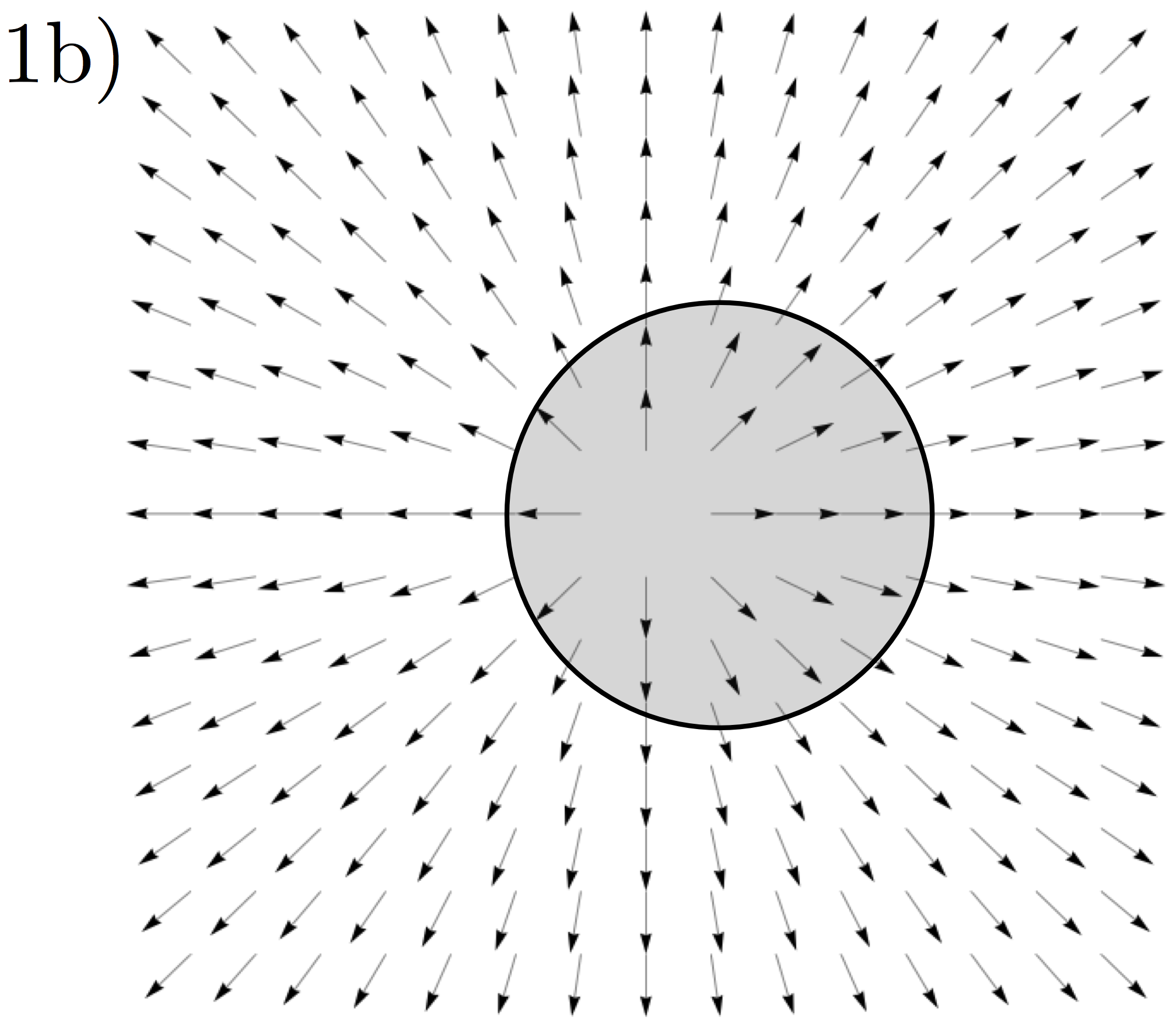} &
\includegraphics[width=0.50\linewidth]{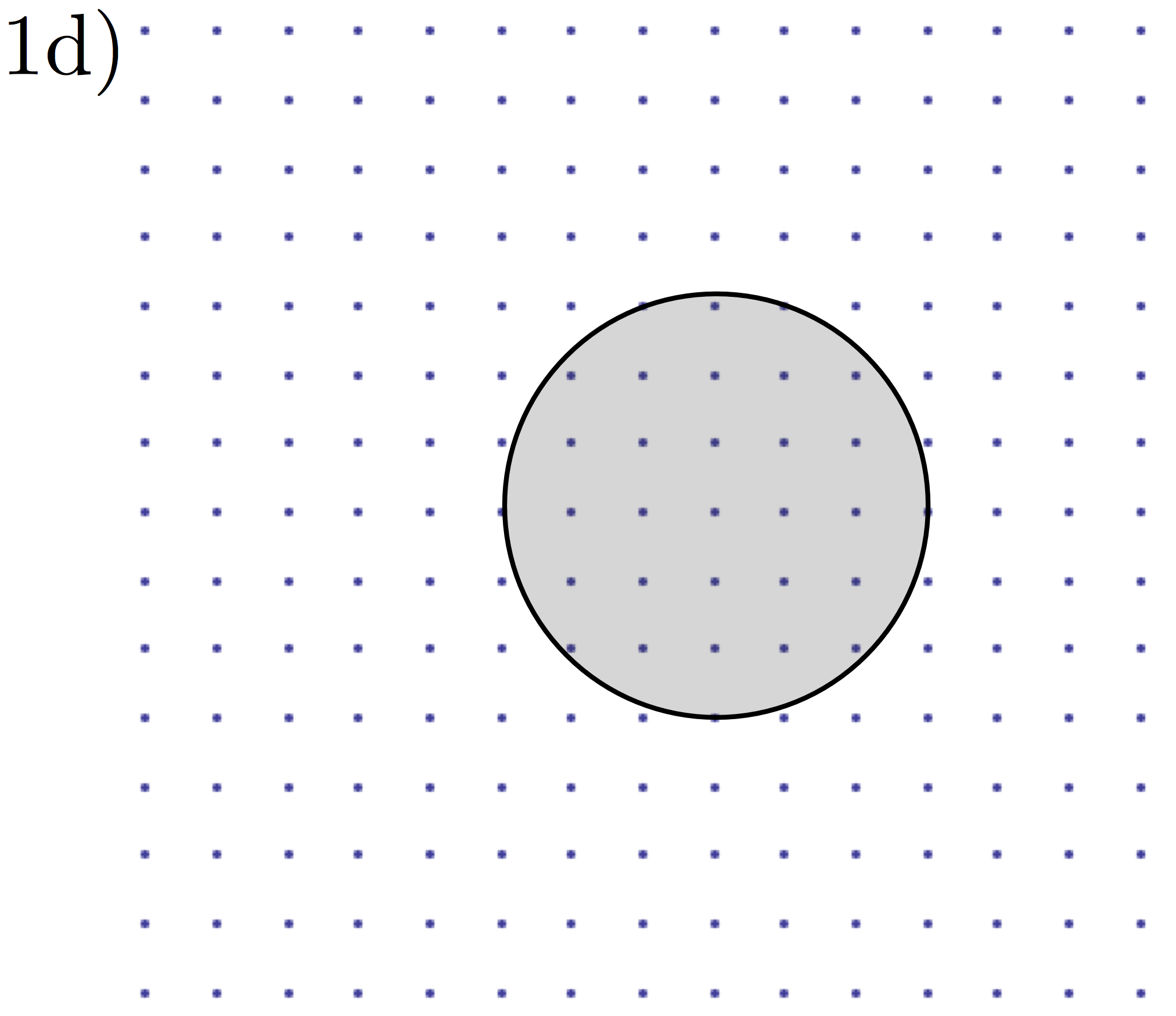}\\
\end{tabular}
\caption{Breakdown of Galilean invariance in graphene. Panel 1a) shows the occupied electronic states in the upper band of graphene in the ground state.  Notice that every state is characterized by a value of momentum (the origin of the arrow) and a pseudospin orientation (the direction of the arrow).  Panel 1b) shows the occupied states {\it after} a Galilean boost.  An observer riding along with the boost would clearly see that the orientation of the pseudospins has changed.  It looks like the pseudospins are subjected to a ``pseudomagnetic field" that causes them to tilt towards the $+{\hat {\bm x}}$ direction.  The appearance of this pseudomagnetic field is the signature of broken Galilean invariance.  In contrast, in a Galilean invariant system [Panels 1c) and 1d)] the energy eigenstates are characterized by momentum only:  an observer riding along with the boost would not see any change in the character of the occupied states.\label{fig:one}}
\end{figure}

The feature of graphene that is ultimately responsible for the large many-body effects on the plasmon dispersion and the Drude weight is {\it broken Galilean invariance}.
The lattice reference frame remains present in the continuum model through the coupling between momenta and pseudospins.  
The oriented  pseudospins provide an ``ether" against which a global boost of the momenta becomes detectable. This is explained in detail in the caption of Fig.~\ref{fig:one}.

Fig.~\ref{fig:one} explains why the plasmon frequency in graphene is so strongly affected by exchange and 
correlation.  In a plasmon mode the region of occupied states (Fermi circle) oscillates back and forth in momentum space under the action of the self-induced electrostatic field.  In graphene, this oscillatory motion is inevitably coupled with an oscillatory motion of the pseudospins.  Since exchange interactions depend on the relative orientation of pseudospins they contribute to plasmon kinetic energy and renormalize the plasmon frequency even at leading order in $q$.

In this article we present a many-body theory of this subtle pseudospin coupling effect and 
discuss the main implications of our findings for theories of charge transport and collective 
excitations in doped graphene sheets. Our manuscript is organized as follows. In Sect.~\ref{sect:theory} we introduce the most important definitions and the basic linear-response functions that control the plasmon dispersion, the Drude weight, and the a.c. conductivity. In Sect.~\ref{sect:diagrammaticPT} we present the approach we have used to calculate the Drude weight and the a.c. conductivity, {\it i.e.} diagrammatic perturbation theory, and the main analytical results.  Identical results can be obtained using a 
kinetic equation approach which will be detailed in a separate publication~\cite{Tse_KE} that is focused on a different application.
Our main numerical results based on this approach are illustrated in Sect.~\ref{sect:drude}. In Sect.~\ref{sect:dvsc} we emphasize how density-density and current-current response functions do not lead to the same results for the Drude weight and a.c. conductivity due to the presence of a rigid cutoff in momentum space. Finally, in Sect.~\ref{sect:discussion} we summarize our findings and draw our main conclusions. Three Appendices (\ref{app:noninteracting}-\ref{app:selfenergy}) highlight some important technical aspects of the diagrammatic calculation, while Appendix~\ref{app:general-continuity-equation} reports on the generalized form of the continuity equation that applies in the presence of a rigid momentum cutoff, and Appendix~\ref{app:thomasfermi} reports numerical results for Thomas-Fermi screened interactions.

\section{Formulation}
\label{sect:theory}

The collective (plasmon) modes of the system described by Hamiltonian ${\hat {\cal H}}$ can be found by solving the following equation$^{\,}$\cite{Giuliani_and_Vignale},
\be\label{eq:plasmon_equation}
1 - V_q {\widetilde \chi}_{\rho\rho}(q,\omega) = 0~,
\ee
where ${\widetilde \chi}_{\rho\rho}(q,\omega)$ is the so-called {\it proper}~\cite{proper_diagrams} 
density-density response function. In the $q \to 0$ limit of interest here we can neglect the distinction between the proper and 
the full causal response function
\be 
\chi_{\rho\rho}(q,\omega) = \frac{{\widetilde \chi}_{\rho\rho}(q,\omega)}{1- V_q {\widetilde \chi}_{\rho\rho}(q,\omega)}.
\ee
We show below that
\be\label{eq:calD}
\lim_{\omega \to 0} \lim_{q \to 0} \Re e~\chi_{\rho\rho}(q,\omega) = 
\frac{{\cal D}}{\pi e^2}~\frac{q^2}{\omega^2}~,
\ee
where ${\cal D}$ is a as yet unidentified density- and coupling-constant-dependent quantity.  
Note the order of limits in Eq.~(\ref{eq:calD}): the limit $\omega \to 0$ is taken in the dynamical sense, {\it i.e.} 
$v q \ll \omega \ll 2\varepsilon_{\rm F}$. Here $\varepsilon_{\rm F}=v k_{\rm F}$ is the Fermi energy and 
$2 \varepsilon_{\rm F}$ is the threshold for vertical inter-band electron-hole excitations. Using Eq.~(\ref{eq:calD}) 
in Eq.~(\ref{eq:plasmon_equation}) and solving for $\omega$ we find that, to leading order in $q$,
\be\label{MainResult}
\omega_{\rm pl}(q\to 0) = \sqrt{\frac{2\pi e^2 n}{\epsilon m_{\rm pl}} q}~,
\ee
where we have introduced the ``plasmon mass", $m_{\rm pl} = \pi e^2 n/{\cal D}$. 

In the dynamical limit, the imaginary-part of the a.c. conductivity, 
$\sigma(\omega) = i e^2 \omega\lim_{q\rightarrow 0} \chi_{\rho\rho}(q,\omega)/q^2$, has the form
\be\label{DrudeWeightForm} 
\Im m~\sigma(\omega) \to \frac{{\cal D}}{\pi \omega}~.
\ee
It then follows from a standard Kramers-Kr\"onig analysis that the real-part of the conductivity has a $\delta$-function peak at $\omega=0$:
$\Re e~\sigma(\omega) = {\cal D} \delta(\omega)$. Thus the quantity ${\cal D}$ introduced in Eq.~(\ref{eq:calD}) 
is the Drude weight. In the presence of disorder the $\delta$-function peak is broadened into a Drude peak, but the Drude weight is preserved for 
weak disorder.

We thus see from Eq.~(\ref{MainResult}) that the Drude weight  completely controls the plasmon dispersion at long wavelengths. 
When electron-electron interactions are neglected ${\cal D}$ tends to the RPA Drude weight 
${\cal D}_0 = 4 \varepsilon_{\rm F} \sigma_0$, where $\sigma_0 = e^2/(4 \hbar)$ is the so-called universal~\cite{kuzmenko_prl_2008,nair_science_2008,wang_science_2008,li_natphys_2008,mak_prl_2008} 
frequency-independent inter-band conductivity of a neutral graphene sheet. In the same limit $m_{\rm pl} \to \hbar k_{\rm F}/v$ and 
\be\label{eq:RPA}
\omega^2_{\rm pl}(q \to 0) = \frac{\varepsilon^2_{\rm F}}{\hbar^2}~\frac{g\alpha_{\rm ee}}{2}~\frac{q}{k_{\rm F}}~,
\ee
where $g=g_{\rm s} g_{\rm v} =4$ is a spin-valley degeneracy factor and we have introduced the dimensionless fine-structure coupling constant 
$\alpha_{\rm ee}= e^2/(\epsilon \hbar v)$, the ratio between the Coulomb energy scale $e^2 k_{\rm F}/\epsilon$ and the kinetic energy scale $\hbar v k_{\rm F}$. 
(The fine-structure constant can be tuned experimentally by changing the dielectric environment surrounding the graphene flake~\cite{jang_prl_2008,mohiuddin_preprint_2008}.) Eq.~(\ref{eq:RPA}) is  the well-known RPA~\cite{wunsch_njp_2006,hwang_prb_2007,polini_prb_2008,principi_prb_2009} 
result for the plasmon dispersion at long wavelengths. 

In the following Section we calculate ${\cal D}$ {\it exactly} to first order in the fine-structure constant $\alpha_{\rm ee}$  by means of diagrammatic perturbation theory, demonstrating in the process that its value is substantially enhanced by electron-electron interactions.  Our results depend on the electron density {\it via} the ultraviolet cutoff $\Lambda = k_{\rm max}/k_{\rm F}$.  Thus, the momentum sums that appear in the evaluation of the diagrams will be restricted in such a way that only single-particle states with wave vectors $k \leq k_{\rm max}$ are involved.
The value of $\Lambda$ varies from $\sim 20$ for a very high-density graphene system with $n \sim 10^{13}~{\rm cm}^{-2}$
to $\sim 100$ for a density $n \sim 5 \times 10^{11}~{\rm cm}^{-2}$ just large enough to screen out the unintended~\cite{yacoby_natphys_2008} inhomogeneities present in samples on substrates. 
We will see that our results are only weakly dependent on $\Lambda$.  If this were not true the Dirac model for electron-electron interactions 
in doped graphene would not be useful and it would be necessary to correctly account for interaction effects at energy scales beyond 
those for which the model is valid.  The cutoff appears only in the well-known electron-electron interaction 
enhancement of the quasiparticle velocity. 

\section{Diagrammatic perturbation theory}
\label{sect:diagrammaticPT}

\begin{figure}
\includegraphics[width=1.0\linewidth]{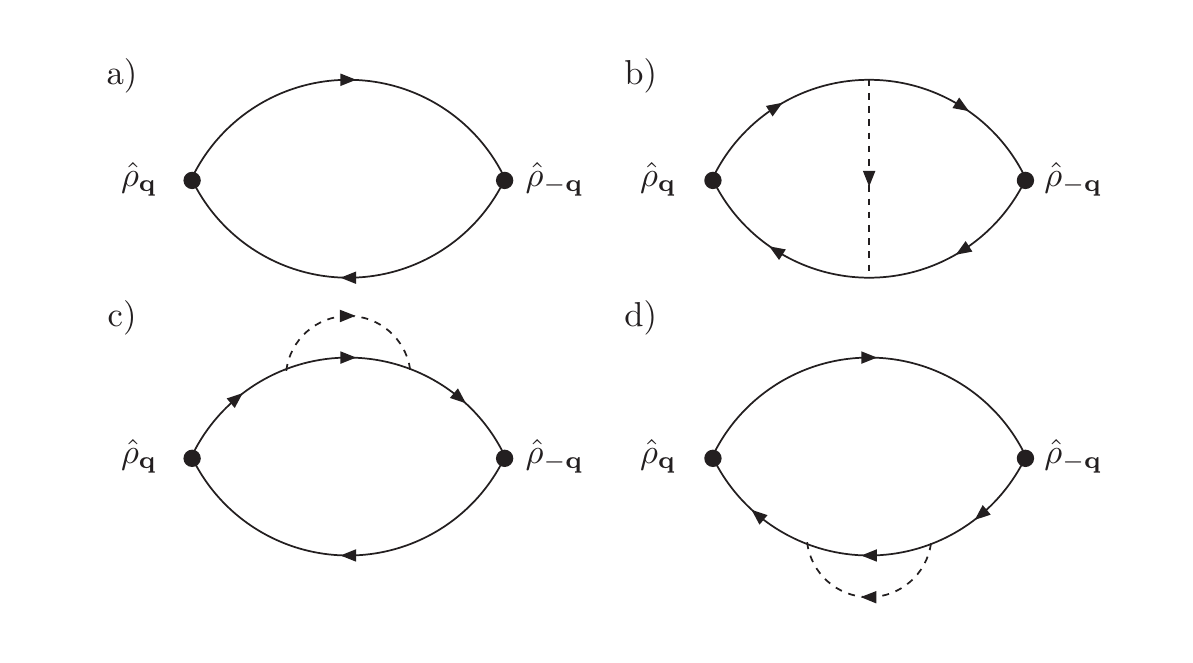} 
\caption{Feynman diagrams for the proper density-density response function $\chi_{\rho\rho}(q,\omega)$ up to first order in the electron-electron interaction. Panel a) The bare bubble diagram. Panel b) Vertex correction. Panels c) and d) Self-energy diagrams.\label{fig:two}}
\end{figure}

In Fig.~\ref{fig:two} we show the diagrams that contribute to the density-density response function $\chi_{\rho\rho}(q,\omega)$ up to first order in the coupling constant; the bare bubble diagram, two self-energy-correction diagrams, and one vertex-correction diagram.  As usual, partial cancellations between the self-energy and vertex corrections play an essential role.
In Fig.~\ref{fig:two} solid lines are noninteracting Green's functions~\cite{mishchenko_prl_2007},
\be\label{G0}
{\bm G}(\kv,\omega)=\frac{1}{2}\sum_{\mu=\pm1} \frac{\openone_\sigma + \mu \sigmav_\kv}{\omega-\xi_{\kv,\mu} +i \eta_{\kv,\mu}}~,
\ee
where $\openone_\sigma$ is the identity matrix in pseudospin space, $\sigmav_\kv = \sigmav\cdot\kv/k$, $\xi_{\kv,\mu}=\mu v k - \varepsilon_{\rm F}$, and $\eta_{\kv,\mu}=\eta~{\rm sgn}(-\xi_{\kv,\mu})$ (with $\eta = 0^+$). Dashed lines are electron-electron interactions.

All wave vectors, $q$, $k$, and $k'$, which appear below are measured in units of $k_{\rm F}$, while frequencies and energies are in units of $2\varepsilon_{\rm F}$.  Below we set $\hbar \to 1$.
We also introduce the density-of-states at the Fermi energy, $\nu(\varepsilon_{\rm F}) = 2\varepsilon_{\rm F}/(\pi v^2)$.

The diagrams in Fig.~\ref{fig:two} are first evaluated in the limit of small $q$ ({\it i.e.} to order $q^2$) and finite $\omega$.  Then, for the real part of $\chi_{\rho\rho}$ we will retain only the terms that scale as $q^2/\omega^2$ for $\omega \to 0$ and thus contribute to the Drude weight, while for the imaginary part of $\chi_{\rho\rho}$, which controls the real part of the a.c. conductivity, we calculate  the full frequency-dependent function. 

Proceeding in this manner we see that the empty bubble in Fig.~\ref{fig:two} ({\it i.e.} the noninteracting diagram), denoted by $\chi_{\rho\rho}^{(0)}(q,\omega)$, reproduces the noninteracting Drude weight ${\cal D}_0$, as expected (see Appendix~\ref{app:noninteracting}): 
\be\label{Rechi0}
\Re e~\chi^{(0)}_{\rho\rho}(q,\omega) =
\frac{{\cal D}_0}{4\pi v^2e^2}~\frac{q^2}{\omega^2} = \frac{1}{8}\nu(\varepsilon_{\rm F})\frac{q^2}{\omega^2}
\ee
and
\be\label{Imchi0}
\Im m~\chi^{(0)}_{\rho\rho}(q,\omega) = -\frac{ \pi}{16 }\nu(\varepsilon_{\rm F})\frac{q^2}{\omega}
\left[\Theta(\omega - 1) + \left\{\omega \to -\omega\right\}\right]~,
\ee 
where $\Theta(x)$ is the usual Heaviside step function and the notation ``$\left\{\omega \to -\omega\right\}$" means that we have to add to the first term in square brackets an identical term in which $\omega$ is interchanged with $-\omega$.

The next diagram is the so-called ``vertex correction", denoted  by $\chi^{({\rm V})}_{\rho\rho}(q,\omega)$, which physically represents the dressing of the external driving field by the internally generated exchange field.  We find that, up to order $q^2$ and for $\omega \to 0$ (see Appendix~\ref{app:vertex} for details),
\ber\label{Rechiex}
\Re e~\chi^{({\rm V})}_{\rho\rho}(q,\omega) &=&
-\frac{1}{32}\alpha_{\rm ee} \nu(\varepsilon_{\rm F}) \frac{q^2}{\omega^2}[V_0(1,1) + 2V_1(1,1)\nonumber\\
&+& V_2(1,1)]
\eer
and
\be\label{Imchiex}
\begin{split}
\Im m~\chi_{\rho\rho}^{({\rm V})}(q,\omega) &= - \frac{\pi}{16} \alpha_{\rm ee} \nu(\varepsilon_{\rm F}) \frac{q^2}{\omega}\\
&\times
\left[ \Theta(\omega-1)J_{\rm V}(\omega,\Lambda) + \left\{\omega \to -\omega\right\}\right]~,
\end{split}
\ee
with
\ber\label{jex}
J_{\rm V}(\omega,\Lambda) &=&\frac{V_2(\omega,1) - V_0(\omega,1)}{2} \nn\\
&+& \frac{\omega}{2}
{\cal P}\int_{1}^{\Lambda}d k\frac{F(\omega,k)}{k^2-\omega^2}
\eer
and $F(\omega,k) = 2 k V_1(\omega,k)+\omega [V_0(\omega,k) + V_2(\omega,k)]$. Here $V_m(k,k')$ are dimensionless Coulomb pseudopotentials~\cite{borghi_ssc_2009},
\be\label{eq:Coulomb_pseudopotentials}
V_m(k,k') = \int_{0}^{2\pi}\frac{d\theta}{2\pi}\frac{\exp{(-i m\theta)}}{q_{\rm TF} + \sqrt{k^2+{k'}^2-2 k k' \cos(\theta)}}~,
\ee
where $q_{\rm TF} = 4 \alpha_{\rm ee}$ is the Thomas-Fermi screening wave vector (in units of $k_{\rm F}$). Making use of these formulas it is easy to check that the vertex correction to the Drude weight [{\it i.e.} Eq.~(\ref{Rechiex})] is {\it negative} and that the integral in the second line of Eq.~(\ref{jex}) converges in the limit $\Lambda \to \infty$.

The last two diagrams are ``self-energy" corrections, denoted  by $\chi_{\rho\rho}^{({\rm SE})}(q,\omega)$, which physically describe the modification of the response function due to exchange energy corrections to the quasiparticle dispersion. We find that (see Appendix~\ref{app:selfenergy} for more details)
\ber\label{RechiSE}
\Re e~\chi^{({\rm SE})}_{\rho\rho}(q,\omega) &=&
\frac{1}{32}\alpha_{\rm ee}\nu(\varepsilon_{\rm F})\frac{q^2}{\omega^2}
[V_0(1,1)+2V_1(1,1) \nn\\
&+&V_2(1,1)] + \frac{1}{32}\alpha_{\rm ee}\nu(\varepsilon_{\rm F})\frac{q^2}{\omega^2} \nn\\
&\times&\int_{1}^{\Lambda} dk~[V_0(1,k) - V_2(1,k)]
\eer
and
\be\label{ImchiSE}
\begin{split}
\Im m~\chi_{\rho\rho}^{({\rm SE})}(q,\omega)& = - \frac{\pi}{16} \alpha_{\rm ee} \nu(\varepsilon_{\rm F}) \frac{q^2}{\omega} \\
&\times
\left[\Theta(\omega-1) J_{\rm SE}(\omega,\Lambda) + \left\{\omega \to -\omega\right\}\right]~,
\end{split}
\ee
where
\be
\left\{
\begin{array}{l}
{\displaystyle J_{\rm SE}(\omega,\Lambda)=\frac{1}{\omega}\Sigma(\omega,\Lambda) - 
\partial_\omega \Sigma(\omega,\Lambda)}\vspace{0.2 cm}\\
{\displaystyle \Sigma(\omega,\Lambda)=\frac{1}{2}\int_1^{\Lambda}dk~k V_1(\omega,k)}
\end{array}
\right.~.
\ee
In writing Eq.~(\ref{ImchiSE}) we have excluded a term proportional to $\delta(\omega-1)$ which is an artifact of perturbation theory
as we explain below.

\section{Drude weight renormalization and a.c. conductivity}
\label{sect:drude}

We now combine the terms calculated in the previous Section.  Extensive cancellations occur between vertex corrections and self-energy contributions. 
For example the first term on the right hand side of Eq.~(\ref{RechiSE}) cancels the vertex contribution (\ref{Rechiex}).  
The final result for the Drude weight to first order in $\alpha_{\rm ee}$ is
\be\label{D1nn}
\frac{\cal D}{{\cal D}_0}=1+\frac{\alpha_{\rm ee}}{4}\int_1^\Lambda dk~[V_0(1,k) - V_2(1,k)]~.
\ee
The real part of the a.c. conductivity is given, to the same order, by
\be\label{sigma1_nn}
\frac{\Re e~\sigma(\omega)}{\sigma_0}=\Theta(\omega-1)\{1+\alpha_{\rm ee}[J_{\rm V}(\omega,\Lambda) + 
J_{\rm SE}(\omega,\Lambda)]\}~.
\ee
Eqs.~(\ref{D1nn})-(\ref{sigma1_nn}) are the most important results of this article. 

\subsection{Long-range interactions}
\label{sect:longrange}

For unscreened Coulomb interactions [$q_{\rm TF}=0$ in Eq.~(\ref{eq:Coulomb_pseudopotentials})] $V_0(1,k)$ decays as $1/k$ at large $k$ (see, for example, Ref.~\onlinecite{borghi_ssc_2009}): we thus find that
\be\label{eq:elegance}
\frac{\cal D}{{\cal D}_0} = \frac{v^\star}{v} + \beta \alpha_{\rm ee}~,
\ee
where
\be\label{eq:velocityenhancement}
\frac{v^\star}{v} = 1+\frac{\alpha_{\rm ee}}{4}\ln{(\Lambda)}
\ee
is the well-known logarithmic velocity enhancement~\cite{gonzalez_nuclearphys_1994} (see also Refs.~\onlinecite{borghi_ssc_2009,Tse_PRB_2007,barlas_prl_2007}) and
\ber\label{eq:betafactor}
\beta &\equiv& \frac{1}{4} \lim_{\Lambda \to \infty} \int_1^\Lambda dk \left[V_0(1,k) - \frac{1}{k} - V_2(1,k)\right] \nn\\
&=&  - \frac{1}{8} + \frac{1}{4\pi}\left(\frac{2}{3} - 4 C\right) + \frac{\ln{4} }{4}\simeq - 0.017~,
\eer
$C \simeq 0.916$ being Catalan's constant. Notice that in Eq.~(\ref{eq:betafactor}) we have taken the limit $\Lambda \to \infty$ since the integrand decays like $1/k^2$ for $k \to \infty$ [$\beta$ reaches its $\Lambda = \infty$ asymptotic value reported in Eq.~(\ref{eq:betafactor}) already at values of $\Lambda$ as small as $\approx 10$]. For all dopings of experimental relevance $\Lambda \gg 1$: in this regime the doping dependence of ${\cal D}$ is logarithmic and stems from the velocity enhancement factor (\ref{eq:velocityenhancement}). The enhancement or suppression of ${\cal D}$ with respect to ${\cal D}_0$ depends on the relative strength of the two terms in Eq.~(\ref{eq:elegance}), which have opposite sign. In the low-density $\Lambda \gg 1$ regime the velocity enhancement completely dominates the ``$\beta$-term" and ${\cal D}/{\cal D}_0 >1$. The enhancement of  ${\cal D}$, or, equivalently of the plasmon frequency, can be understood qualitatively by noting that electron-electron interactions 
{\it reduce} the pseudospin susceptibility~\cite{polini_unpublished} and therefore {\it increase} the pseudospin stiffness, {\it i.e.} the energy that is required to align pseudospins along a given direction.  
According to the discussion given in the Introduction, the larger pseudospin stiffness results in higher energy of plasma oscillations.  
This observation is consistent with the fact that in a Landau Fermi liquid description~\cite{polini_unpublished} the suppression of the pseudospin susceptibility would be driven by the many-body enhancement of the density of states factor  $v^\star/v$, while interactions between the quasiparticles produce the opposite effect.  Once again we must conclude that the many-body enhancement of the plasmon frequency is intimately connected to the many-body enhancement of  the quasiparticle velocity $v^\star/v$.  Any physical mechanism that {\it reduces} $v^\star/v$ without affecting the interaction between quasiparticles could in principle result in a reduction of the plasmon frequency and Drude weight.

\begin{figure}
\includegraphics[width=1.0\linewidth]{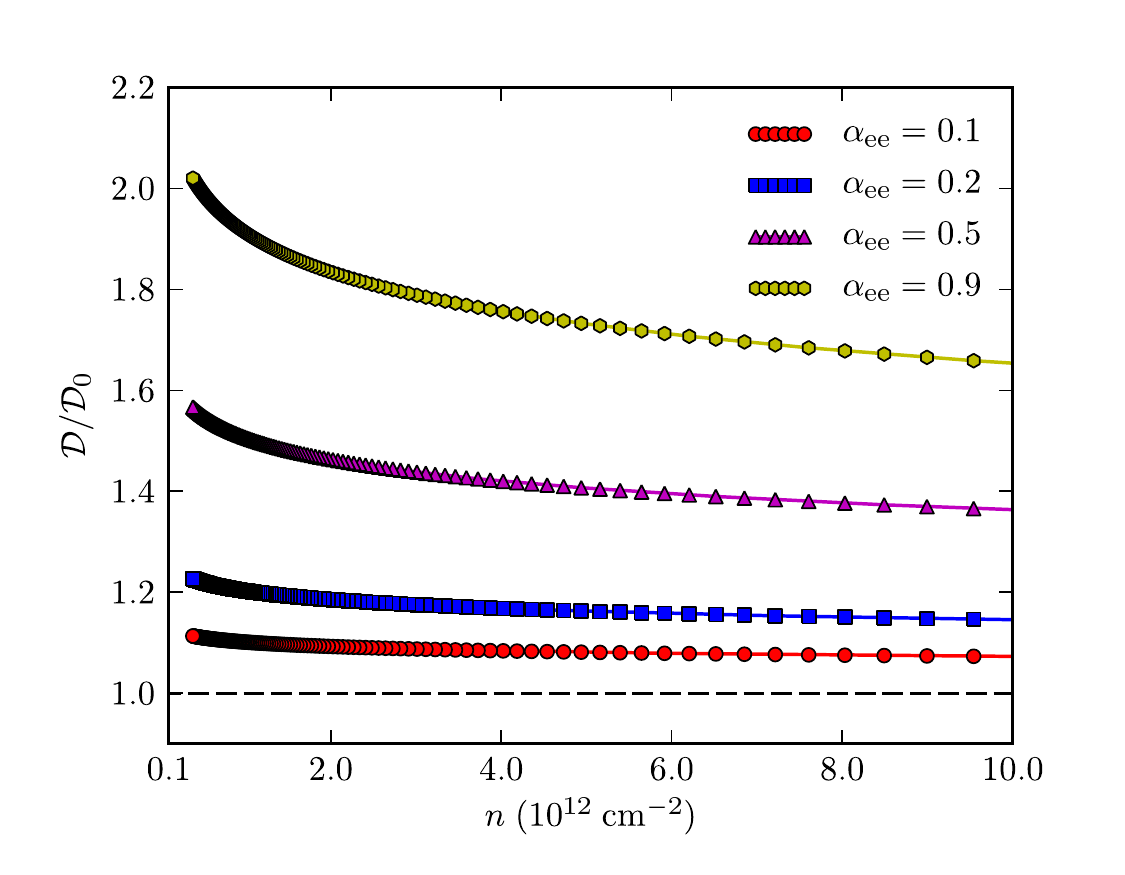} 
\caption{(Color online) The ratio ${\cal D}/{\cal D}_0$ between the interacting value of the Drude weight ${\cal D}$, 
calculated from Eq.~(\ref{D1nn}), and the RPA value ${\cal D}_0 = 4\varepsilon_{\rm F}\sigma_0$ is plotted as a function of electron density $n$ (in units of $10^{12}~{\rm cm}^{-2}$) for different values of graphene's fine-structure constant $\alpha_{\rm ee}$. The value $\alpha_{\rm ee} = 0.9$ is believed to be appropriate for graphene deposited on ${\rm SiO}_2$, the other side exposed to air. Note that ${\cal D}/{\cal D}_0 >1$ and that it depends weakly on carrier density. \label{fig:three}}
\end{figure}

\begin{figure}
\includegraphics[width=1.0\linewidth]{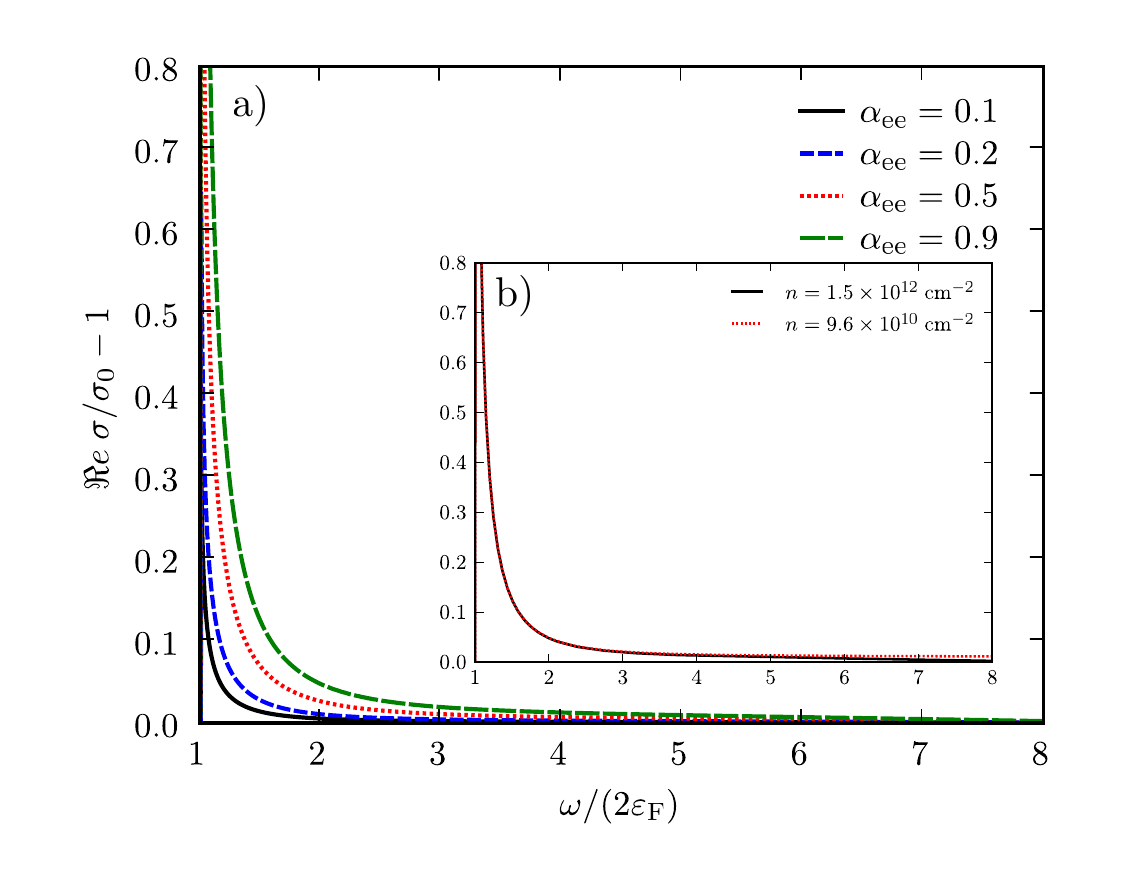} 
\caption{(Color online) Panel a) Deviation of the real part of the a.c. conductivity $\Re e~\sigma(\omega)$ from the noninteracting universal value $\sigma_0=e^2/(4\hbar)$, as a function of frequency $\omega/(2\varepsilon_{\rm F}$), calculated from Eq.~(\ref{sigma1_nn}) 
for several values of the fine structure constant $\alpha_{\rm ee}$ and for $n = 1.5 \times 10^{12}~{\rm cm}^{-2}$ ($\Lambda =50$). 
Panel b) Same as in the main panel but for a fixed value of the fine structure constant ($\alpha_{\rm ee} = 0.9$) and two different values of doping (corresponding to $\Lambda =50$ and $100$). Notice that the dependence of the conductivity on the value of the cutoff $\Lambda$ is almost invisible at low frequencies and becomes visible only at frequencies several times $\varepsilon_{\rm F}$.\label{fig:four}}
\end{figure}

The Drude weight ${\cal D}$ and the real part of the a.c. conductivity $\sigma(\omega)$ are plotted as functions of doping $n$ and $\omega$ respectively in Figs.~\ref{fig:three} and~\ref{fig:four}. 
We observe that the Drude weight (and hence the coefficient of $q^{1/2}$ in the plasmon dispersion relation) is substantially {\it enhanced} above the noninteracting value.  As we showed above in Eqs.~(\ref{eq:elegance})-(\ref{eq:velocityenhancement}), the enhancement grows slowly (logarithmically) as a function of $\Lambda$ [it grows {\it linearly} with $\Lambda$ for short-range interactions -- see Eq.~(\ref{Drude_Weight_SR}) in Sect.~\ref{sect:shortrange}]. 
The a.c. conductivity is likewise enhanced above the threshold $\omega=2\varepsilon_{\rm F}$. 

As we have mentioned above, an unphysical term proportional to $\delta(\omega-1)$ has been omitted from Eq.~(\ref{sigma1_nn}). 
This singular contribution is due to the shift of the inter-band absorption threshold from the bare value $2 \varepsilon_{\rm F}$ to the dressed value $2 \varepsilon_{\rm F}^\star$. In first-order perturbation theory the absorption shift appears as a $\delta(\omega-1)$ contribution to the integrand for frequency integrals.  

Fig.~\ref{fig:four} shows that for $\omega \gg 2\varepsilon_{\rm F}$ but $\omega\ll 2\Lambda \varepsilon_{\rm F}$ the a.c. conductivity approaches the high-frequency universal value $\sigma_0$.  However, as $\omega$ becomes comparable to  $2\Lambda \varepsilon_{\rm F}$ the conductivity decreases and, in fact, has an unphysical logarithmic divergence at $\omega=2\Lambda \varepsilon_{\rm F}$.  This makes perfect sense, since our model is only valid for energies that are much smaller that the cutoff energy $2\Lambda \varepsilon_{\rm F}$.  As long as this condition is met, the calculated spectrum is essentially independent of the cutoff.  This is a very satisfactory feature of the present calculation.

\subsection{Short-range interactions}
\label{sect:shortrange}

It is instructive to examine how the results presented in the main body of this Section change if the electron-electron interaction 
is assumed to be of ultra-short range in space, {\it i.e.} $V_0 = v = {\rm const}$ and all other moments $V_m$ with $m \geq 1$ are zero. 
The calculations can be carried out in a completely analytical fashion with the following results:
\be\label{Drude_Weight_SR}
\left.\frac{\cal D}{{\cal D}_0}\right|_{\rm sr} =1+\frac{\alpha_{\rm ee} {\bar v}}{4}(\Lambda-1)~,
\ee
where ${\bar v} = \epsilon k_{\rm F} v/(2\pi e^2)$, and
\be\label{Optical_Conductivity_LR}
\left.\frac{\Re e~\sigma(\omega)}{\sigma_0}\right|_{\rm sr} = \Theta(\omega-1)[1+\alpha_{\rm ee}J_{\rm sr}(\omega,\Lambda)]~, 
\ee
where
\be
J_{\rm sr}(\omega,\Lambda) = \frac{{\bar v}}{2}\left\{-1+\frac{\omega}{2}\ln\left[\frac{(\omega+1)(\Lambda-\omega)}{(\omega-1)(\Lambda+\omega)}\right]\right\}~.
\ee
Note that $J_{\rm sr}(\omega,\Lambda)$ becomes independent of $\Lambda$ for $\Lambda \to \infty$. It is easy to see that, in this limit, 
$\Re e~\sigma(\omega)$ approaches the universal value for large $\omega$, since $J_{\rm sr}(\omega,\infty)$ goes to zero like 
${\bar v}/(6\omega^2)$ for $\omega \to \infty$.  If, on the other hand, $\omega$ is allowed to tend to infinity {\it before} $\Lambda$ then unphysical cutoff-related features appear, such as a logarithmic divergence at $\omega = \Lambda$.

Let us make a brief comment on the limit of zero doping of our theory. In this limit we predict that the strength of the Drude peak vanishes as the absorption threshold at $2\varepsilon_{\rm F}$ moves toward zero. Then for finite but arbitrarily small $\omega$ the optical conductivity approaches the universal value $\sigma_0$, yielding a result that is independent of $\alpha_{\rm ee}$.

\section{Density response {\it versus} current response}
\label{sect:dvsc}

In many models of electronic systems, gauge invariance and the continuity equation allow us to express the density-density response function in terms of the current-current response function.  In the present model the relation would take the form 
\be\label{contin}
\chi_{\rho\rho}(q,\omega) = \frac{v^2q^2}{\omega^2}\chi_{\sigma\sigma}(q,\omega) + 
\frac{vq}{\omega^2} \langle[{\hat \sigma}_\qv, {\hat \rho}_{-\qv}]\rangle~,
\ee
where ${\hat \sigma}_\qv$ is the {\it longitudinal} component (parallel to $\qv$) of the pseudospin-density fluctuation and 
$[{\hat \sigma}_\qv, {\hat \rho}_{-\qv}]$ is an {\it anomalous} commutator~\cite{sabio_prb_2008}, reminiscent of the commutator of Fourier-component-resolved density fluctuations in a 1D Luttinger liquid~\cite{giamarchi_book}. Because the current in the Dirac model is proportional to the pseudospin density,  $\chi_{\sigma\sigma}(q,\omega)$ in Eq.~(\ref{contin}) is the current-current response function.
Eq.~(\ref{contin}) works perfectly at the noninteracting level, but {\it fails} when electron-electron interactions are taken into account. 
Due to the cutoff in momentum space, it turns out that the continuity equation,
\be\label{eq:continuity}
i\partial_t {\hat \rho}_\qv =  \qv \cdot {\hat {\bm j}}_\qv~,
\ee
with the current-density operator ${\hat {\bm j}}_\qv$ given by
\be\label{eq:current_pseudospin_relation}
{\hat {\bm j}}_\qv = v \sum_{\kv, \alpha, \beta} {\hat \psi}^\dagger_{\kv - \qv, \alpha} \sigmav_{\alpha\beta} {\hat \psi}_{\kv, \beta}~,
\ee
is no longer satisfied in the interacting system (see Appendix~\ref{app:general-continuity-equation}). 

Similar conclusions have been reached by Mishchenko~\cite{mishchenko_epl_2008} in the context of calculations of the a.c. conductivity of {\it undoped} graphene sheets. This topic has indeed attracted considerable interest~\cite{sheehy_prl_2007,herbut_prl_2008,mishchenko_epl_2008, katsnelson_epl_2008, sheehy_prb_2009} and has been at the center of a dispute.~\cite{herbut_prl_2008,mishchenko_epl_2008}  Mishchenko\cite{mishchenko_epl_2008}, in particular, was the first to clarify that the calculation of $\sigma(\omega)$ based on the density-density response function predicts a very small correction over the noninteracting conductivity, which is also in agreement with the experimental findings~\cite{kuzmenko_prl_2008,nair_science_2008}, while methods based on the current-current response function~\cite{herbut_prl_2008} or on the kinetic equation~\cite{mishchenko_epl_2008} predict very different results. 
We have also found~\cite{polini_arxiv_2009} that a naive application of Eq.~(\ref{contin}) to the interacting system produces strongly cutoff-dependent results for both the Drude weight and the a.c. conductivity of the {\it doped} system. Moreover, it yields a qualitatively different behavior of the Drude weight with respect to that found in the present article [see Eq.~(\ref{D1nn}) and Fig.~\ref{fig:three}]. The Drude weight calculated from the current-current response is {\it suppressed}~\cite{polini_arxiv_2009} rather than enhanced.  On the other hand, if Eq.~(\ref{contin}) is corrected to take into account the modification of the continuity equation due to e-e interactions, then the results of the present analysis are recovered.  The conclusion is that it is generally safer in an effective low-energy theory with a rigid cutoff to work with the density-density response function $\chi_{\rho\rho}(q,\omega)$ rather than with the current-current response function, since the calculation of the former places less weight on high-energy intermediate states, which are not properly described.  This observation is consistent with the findings of other authors~\cite{mishchenko_epl_2008}. An alternative solution is to use a modified Coulomb interaction~\cite{mishchenko_epl_2008,sheehy_prb_2009} such that the continuity equation (\ref{contin}) remains valid even in the interacting system.

\section{Discussion and conclusions}
\label{sect:discussion}

Large Fermi velocity enhancements due to exchange interactions, like the velocity enhancement that occurs in graphene,
are common in the theory of solids.  Often the role of velocity enhancements are fully cancelled in response functions by vertex corrections.
For example the Fermi velocity of an ordinary two-dimensional electron gas diverges when screening is neglected, but the 
corresponding reduction in density-density response is absent when vertex corrections like those which appear in this paper are included.
From this point of view the main finding of this paper, supported by an explicit first order perturbation theory calculation, is that 
no such cancellation occurs for graphene's well-known velocity enhancement.  The difference is easy to understand. 
In the case of an ordinary two-dimensional electron gas the velocity enhancement is due to the rapid variation of the unscreened exchange self-energy as one goes from occupied states to empty states across the Fermi surface. 
When the density changes, the energy at which the velocity peak occurs also changes, 
negating its influence on response.  In perturbation theory, this effect is captured by the vertex correction.
In graphene on the other hand, the velocity enhancement occurs over a broad range of 
energies centered on the Dirac point, not the Fermi surface.  Density response is influenced by enhanced 
velocities which increase the energy cost of changing the electron density.  In ordinary two-dimensional electron gases
the logarithmic velocity enhancement vanishes in any event once screening is accounted for.  In graphene, on the other 
hand the enhancement comes from interactions at wavevectors much larger than the Fermi energy at which only 
inter-band screening, which does not change the long range $1/r$ behavior, is relevant.  Neither screening 
nor vertex corrections fully counter the enhanced Drude weight due to graphene's Dirac point velocity enhancement.

We note that a recent experiment~\cite{Berkeley} has clearly established the presence of a strong Drude peak which 
develops in graphene as the carrier density is increased.  These authors conclude that the Drude weight is reduced 
compared to the noninteracting electron theory, instead of being increased as predicted by this theoretical analysis.
It will be interesting to see whether or not this experimental conclusion changes as sample quality improves and 
it becomes possible to more cleanly separate inter-band and Drude conductivity contributions over a wider 
regime of carrier density.

\begin{figure}
\includegraphics[width=1.0\linewidth]{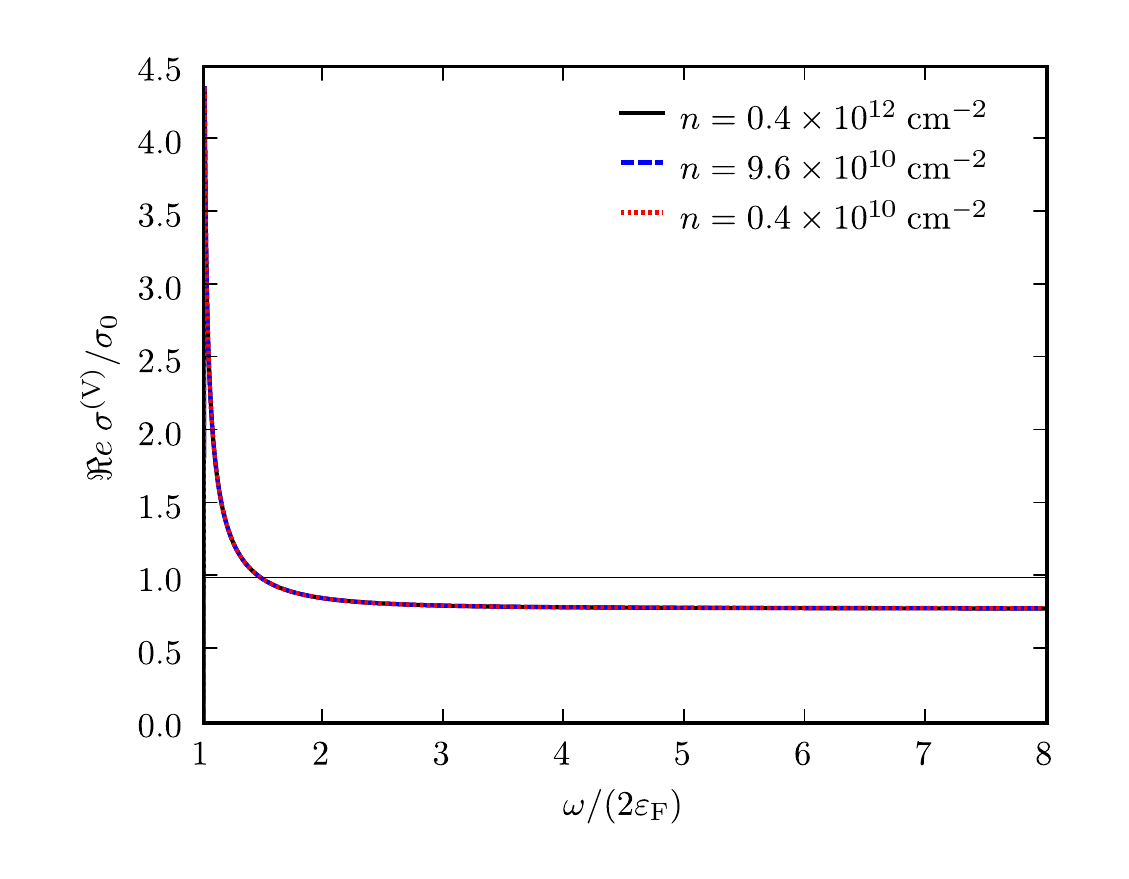} 
\caption{(Color online) The real part of the a.c. conductivity (in units of $\sigma_0$) as a function of $\omega/(2\varepsilon_{\rm F})$ calculated by including {\it only} the vertex correction [{\it i.e.} by neglecting self-energy diagrams in panels c) and d) in Fig.~\ref{fig:two}] for $\alpha_{\rm ee} = 0.9$ and different values of doping $n$ (corresponding to $\Lambda = 100, 200$, and $1000$).  These data have been calculated using long-range (unscreened) interactions, {\it i.e.} $q_{\rm TF} = 0$ in Eq.~(\ref{eq:Coulomb_pseudopotentials}). We remind the reader that, in the absence of disorder and within first-order perturbation theory, $\Re e~\sigma(\omega) = 0$ for $\omega < 2\varepsilon_{\rm F}$.
Notice that $\Re e~\sigma^{({\rm V})}(\omega)$ converges rapidly in the limit $\Lambda \to \infty$ [since, as already stated in the main text, the integral in the second line of Eq.~(\ref{jex}) converges in the same limit] and that it drops well {\it below} the universal value $\sigma_0$ at large $\omega$. \label{fig:five}}
\end{figure}

Before concluding we would like to mention that a new theoretical paper reporting a study of the effect of electron-electron interactions
on  the conductivity of doped graphene steets~\cite{peres_prl_2010} has appeared recently. (We refer the reader to Ref.~\onlinecite{theoryworkopticalconductivity} for studies of band-structure, disorder, phonon, and strain effects.) A quantitative comparison between our findings and theirs is not possible since their study takes disorder into account while our study is for clean graphene. However, we would like to stress that the authors of this paper have treated interactions only at the vertex level neglecting self-energy effects. 
Even though it is well known that this approximation is not ``conserving"~\cite{kadanoff-baym} ({\it e.g.} breaks gauge-invariance) even in a standard parabolic-band electron liquid, we can adopt the same strategy in our calculations and artificially ``switch-off" self-energy contributions to $\Re e~\sigma(\omega)$ [diagrams in panels c) and d) in Fig.~\ref{fig:two}].  
If this procedure is followed, we find the results shown in Fig.~\ref{fig:five}: it is evident from this point that the neglect of self-energy insertions 
is responsible for the reduction of $\Re e~\sigma(\omega)$ below $\sigma_0$ at large frequencies found in Ref.~\onlinecite{peres_prl_2010}.

Finally, let us comment on the broader implications of our results. Effects similar to those described in this article 
occur in graphene bilayers~\cite{tse_prb_2009} and are also expected in other few-layer systems. 
The lack of Galilean invariance also affects the cyclotron resonance frequency when the 2D sheet of graphene is placed in a perpendicular magnetic field~\cite{jiang_prl_2007,deacon_prb_2007,henriksen_prl_2010,iyengar_prb_2007,bychkov_prb_2008,roldan_prb_2009} since Kohn's theorem~\cite{kohn_pr_1961}, which asserts the absence of many-body effects in cyclotron resonance, is not applicable in this case. (The impact of broken Galilean invariance on the collective cyclotron motion in graphene has been studied in Ref.~\onlinecite{mueller_prb_2008} in the high-temperature hydrodynamic regime in which Landau levels are not well resolved.) Undoubtedly much interesting physics, potentially useful for applications in opto-electronics, has still to be learned from the study of graphene and other~\cite{agarwal_arXiv_2010} non-Galilean invariant systems.

\acknowledgments

S.H.A. and G.V. acknowledge support from NSF Grant No. 0705460.
M.P. acknowledges financial support by the 2009/2010 CNR-CSIC scientific cooperation project.
Work in Austin was supported by the Welch Foundation and by SWAN.
S.H.A. acknowledges support and the kind hospitality of the IPM in Tehran, Iran, during the final stages of this work.
We thank Igor Herbut for useful comments and clarifications about the theory of the undoped system.

\appendix
\section{Calculation of the noninteracting contribution to the dynamical density-density response function}
\label{app:noninteracting}
The noninteracting response function $\chi^{(0)}_{\rho\rho}(q,\omega)$, {\it i.e.} the empty bubble diagram in Fig.~\ref{fig:two}a), reads
\ber\label{chin0}
\chi^{(0)}_{\rho\rho}(q,\omega)
&=& \frac{1}{S}\sum_{\kv}\int_{-\infty}^{+\infty}
\frac{d\epsilon}{2\pi i}{\rm Tr}\left[{\bm G}(\kv_-,\epsilon){\bm G}(\kv_+,\epsilon+\omega)\right] \nn\\
&=& 2 \int\frac{d^2{\bm k}}{(2\pi)^2}\sum_{\mu,\nu}\frac{n_{\kv_-,\mu} - n_{\kv_+,\nu}}{\omega + \xi_{\kv_-,\mu} - \xi_{\kv_+,\nu} + i \eta} \nn\\
&\times&{\cal F}_{\mu\nu}(\kv_-,\kv_+)~.
\eer
Here  $\kv_\pm=\kv\pm\qv/2$ and  ${\rm Tr} = g {\rm Tr}_\sigma$, where ${\rm Tr}_\sigma$ is the trace over pseudospin degrees-of-freedom, and ${\cal F}_{\mu \nu}(\kv,\kv') = 1+\mu\nu\cos{(\phi_{\kv} - \phi_{\kv'})}$, $\phi_\kv$ being the angle between $\kv$ and the ${\hat {\bm x}}$ axis.

We are interested in the long-wavelength $q\to 0$ limit of the density-density response function: we thus 
expand Eq.~(\ref{chin0}) to second order in $q$:
\be\label{cosxpan}
{\cal F}_{\mu\nu}(\kv_-,\kv_+)=(1+\mu\nu)-\mu\nu\frac{q^2\sin^2(\phi_\kv)}{2k^2}~,
\ee
where we have assumed that $\qv$ is along the ${\hat {\bm x}}$ ({\it i.e.} that $\phi_\qv=0$).
When $\mu=\nu=-1$, the ``Lindhard ratio" in Eq.~(\ref{chin0}) vanishes. For $\mu = -\nu$, Eq.~(\ref{cosxpan}) is already of order 
$q^2$ and one can just put $q = 0$ in the Lindhard ratio. For $\mu=\nu=+1$ we need instead to expand the Lindhard ratio 
to second order in $q$ with the result
\ber\label{lindxpan}
\frac{n_{\kv_-, +} - n_{\kv_+, +}}{\omega + \xi_{\kv_-, +} - \xi_{\kv_+, +} + i \eta}
&= & \frac{q\cos(\phi_\kv)~\delta(k_{\rm F}-k)}{\omega}\nn\\
&+&\frac{v q^2\cos^2(\phi_\kv)\delta\left(k_{\rm F}-k\right)}{\omega^2}~.\nn\\
\eer
Here we have used $n_{\kv_\pm, +} \simeq n_{\kv, +} \mp q\cos(\phi_{\kv})\delta(k_{{\rm F}}-k)/2$, and we have dropped 
$i\eta$ in the denominator since the expansion on the right hand side is always real. 
The first term in Eq.~(\ref{lindxpan}) gives zero contribution to the response function after angular integration. 
Now, replacing Eqs.~(\ref{cosxpan}) and~(\ref{lindxpan}) in Eq.~(\ref{chin0}), we find
\ber\label{chinn0_q2}
\chi_{\rho\rho}^{(0)}(q,\omega)&=&
\frac{4v q^2}{\omega^2}\int\frac{d^2 \kv}{(2\pi)^2}\cos^2(\phi_\kv)\delta(k_{\rm F}-k) \nn\\
&+&q^2\sum_{\mu}\int\frac{d^2 \kv}{(2\pi)^2}\frac{\sin^2(\phi_\kv)}{k^2}\frac{n_{\kv,\mu} - n_{\kv, {\bar \mu}}}{\omega + 2\mu v k+i \eta}~, \nn\\
\eer
where ${\bar \mu} = -\mu$. Performing the integral over $\kv$ and introducing dimensionless variables, one finds Eqs.~(\ref{Rechi0}) and~(\ref{Imchi0}) for the real and imaginary parts of the noninteracting response function, respectively.

\section{Calculation of the vertex correction}
\label{app:vertex}

The vertex correction [{\it i.e.} see Fig.~\ref{fig:two}b)] contribution to the density-density response function reads
\begin{widetext}
\ber\label{chin_ex}
\chi^{({\rm V})}_{\rho\rho}(q,\omega)&=& - \frac{1}{S^2} 
\sum_{\kv,\kv'} V_{\kv-\kv'}\int_{-\infty}^{+\infty}\frac{d \epsilon}{2\pi i}\int_{-\infty}^{+\infty}\frac{d\epsilon'}{2\pi i}
{\rm Tr}\left[{\bm G}(\kv_-,\epsilon){\bm G}(\kv_+,\epsilon+\omega){\bm G}(\kv'_+,\epsilon'+\omega){\bm G}(\kv'_-,\epsilon')\right]\nn\\
& = & -\frac{1}{16}\int \frac{d^2 \kv}{(2\pi)^2}\int \frac{d^2 \kv'}{(2\pi)^2}\sum_{\mu,\nu, \mu',\nu'}V_{\kv-\kv'}
\frac{n_{\kv_-, \mu} - n_{\kv_+, \nu}}{\omega + \xi_{\kv_-,\mu} - \xi_{\kv_+,\nu} + i\eta}
\frac{n_{\kv'_-, \mu'} - n_{\kv'_+, \nu'}}{\omega + \xi_{\kv'_-, \mu'} - \xi_{\kv'_+, \nu'} + i\eta}\nn\\
&\times &
{\rm Tr}[
(\openone_\sigma + \mu \sigmav_{\kv_-})
(\openone_\sigma + \nu \sigmav_{\kv_+})
(\openone_\sigma + \nu' \sigmav_{\kv'_+})
(\openone_\sigma + \mu' \sigmav_{\kv'_-})
]~.
\eer
\end{widetext}
The trace in the previous equation can be expanded in powers of $q$:
\be
\begin{split}
&{\rm Tr}[
(\openone_\sigma + \mu \sigmav_{\kv_-})
(\openone_\sigma + \nu \sigmav_{\kv_+})
(\openone_\sigma + \nu' \sigmav_{\kv'_+})
(\openone_\sigma + \mu' \sigmav_{\kv'_-})
]\\
&=16\sum_{n=0}^{\infty} w_{n}\left(\kv,\kv',\phi_\qv;\mu,\nu,\mu',\nu'\right)q^n~.
\end{split}
\ee
If we choose $\qv=q {\hat {\bm x}}$, it is straightforward to show that
\be\label{eq:w0}
w_0=\frac{(1+\mu\nu)(1+\mu'\nu')}{2}{\cal F}_{\mu\mu'}(\kv,\kv')~,
\ee
and
\be\label{eq:w1}
w_1=(\mu-\nu)(\mu'+\nu') \frac{\sin(\phi_\kv)\sin(\phi_{\kv'}-\phi_{\kv})}{4 k} + {\rm Perm}~,
\ee
where ``${\rm Perm}$" is obtained from the first term in Eq.~(\ref{eq:w1}) by interchanging primed with non-primed variables. 
The complete expression of $w_2$ is quite cumbersome and will not be reported here.  
The only terms that contribute to $\chi^{({\rm V})}_{\rho\rho}(q,\omega)$ up to order $q^2$ are given by
\be
\left. w_2\right|_{\nu={\bar \mu},\nu'={\bar \mu}'}=
\frac{\sin(\phi_{\kv})\sin(\phi_{\kv'})}{2 k k'}{\cal F}_{\mu\mu'}(\kv, \kv')~.
\ee
Collecting all terms up to ${\cal O}(q^2)$ we can write $\chi^{({\rm V})}_{\rho\rho}(q,\omega)$ in the following form:
\be\label{chin_ex_q2}
\chi^{({\rm V})}_{\rho\rho}(q,\omega) =
\sum_{n=0}^{2}\chi^{({\rm V}-n)}_{\rho\rho}(q,\omega)~,
\ee
where $\chi^{({\rm V}-n)}_{\rho\rho}(q,\omega)$ denotes terms proportional to $w_n$. We find that
\begin{widetext}
\ber\label{chin_ex_w0}
\chi^{({\rm V}-0)}_{\rho\rho}(q,\omega)& =& -2 \int \frac{d^2 \kv}{(2\pi)^2}\int \frac{d^2 \kv'}{(2\pi)^2}
~V_{\kv-\kv'}\left[1+\cos\left(\phi_{\kv'}-\phi_{\kv}\right)\right]
\frac{n_{\kv_-, +} - n_{\kv_+, +}}{\omega + v|\kv_-| - v|\kv_+| + i\eta}
\frac{n_{\kv'_-, +} - n_{\kv'_+, +}}{\omega + v|\kv'_-| - v|\kv'_+| + i\eta}\nn \\
& = &-\frac{2q^2}{\omega^2}\int \frac{d^2 \kv}{(2\pi)^2}\int \frac{d^2 \kv'}{(2\pi)^2}v_{\kv-\kv'}
\delta(k_{\rm F}-k)\delta(k_{\rm F}-k')
\cos(\phi_{\kv})\cos(\phi_{\kv'})\left[1+\cos(\phi_{\kv'}-\phi_{\kv})\right]~.
\eer
%\end{widetext}
%
Introducing the Coulomb pseudopotentials $V_m$, performing the integrations over ${\bm k}$ and ${\bm k}'$, and using dimensionless variable we find immediately Eq.~(\ref{Rechiex}). 

For $n=1$ we find
%
%\begin{widetext}
\ber\label{chin_ex_w1}
\chi^{({\rm V}-1)}_{\rho\rho}(q,\omega) &=& -\frac{2q^2}{\omega}\sum_{\mu}
\int \frac{d^2 \kv}{(2\pi)^2}\int \frac{d^2 \kv'}{(2\pi)^2}\frac{V_{\kv-\kv'}}{k}
\delta(k_{\rm F}-k')\sin(\phi_{\kv})\cos(\phi_{\kv'}) \sin(\phi_{\kv'}-\phi_{\kv}) \frac{\mu \left(n_{\kv,\mu}-n_{\kv,{\bar \mu}}\right)}{\omega+2\mu v k + i\eta} \nn\\
& = & -\frac{q^2k_{\rm F}}{8\pi^2\omega}\sum_{\mu} {\cal P}\int_{k_{\rm F}}^{\infty}dk~\frac{V_0(k,k_{{\rm F}})-V_2(k,k_{{\rm F}})}{\omega+2\mu v k+i\eta}~.
\eer
%\end{widetext}
%
Since this contribution vanishes for $\omega \to 0$ it does not contribute to the renormalization of the Drude weight. Nevertheless, its imaginary part for finite positive frequency gives precisely the term in the first line of Eq.~(\ref{jex}).

Finally, for $n=2$, we find that
%
%\begin{widetext}
\ber\label{chin_ex_w2}
\chi^{({\rm V}-2)}_{\rho\rho}(q, \omega) &=& -\frac{q^2}{2}\sum_{\mu, \mu'}\mu\mu'
\int \frac{d^2 \kv}{(2\pi)^2}\int \frac{d^2 \kv'}{(2\pi)^2} \frac{V_{\kv-\kv'}}{kk'}
\sin(\phi_{\kv})\sin(\phi_{\kv'})\frac{\left(1-n_{\kv,+}\right)\left(1-n_{\kv',+}\right)}{\left(\omega + 2\mu v k +i \eta\right)
\left(\omega + 2\mu' v k' + i \eta\right)}{\cal F}_{\mu\mu'}(\kv,\kv') \nn\\
& = & -\frac{q^2}{2^5\pi^2}\sum_{\mu,\mu'}{\cal P}\int_{k_{\rm F}}^{\infty}dk~\int_{k_{\rm F}}^{\infty}dk'~
\frac{V_0(k,k')+V_2(k,k')+2\mu \mu' V_1(k,k')}{(\omega+2\mu v k+i \eta)(\omega+2\mu' v k'+i \eta)}~.
\eer
\end{widetext}
It is possible to show that this expression does not scale like $\omega^{-2}$ for $\omega\to 0$: thus 
it does not contribute to the renormalization of the Drude weight. The imaginary part of Eq.~(\ref{chin_ex_w2}) at finite frequency gives precisely the term in the second line of Eq.~(\ref{jex}). Summing the contributions to $\Im m~\chi^{({\rm V})}_{\rho\rho}(q,\omega)$ coming from the two terms with $n=1$ and $n=2$ we find Eq.~(\ref{Imchiex}).

\section{Calculation of the self-energy insertions}
\label{app:selfenergy}

We now turn to calculate the two first-order self-energy diagrams in Figs.~\ref{fig:two}c) and~d). 
The first diagram reads
\begin{widetext}
\ber\label{chin_se1}
\chi^{({\rm SE}-{\rm c})}_{\rho\rho}(q,\omega)
& = & - \int \frac{d^2 \kv}{(2\pi)^2}\int \frac{d^2 \kv'}{(2\pi)^2} V_{\kv-\kv'}\int_{-\infty}^{+\infty}\frac{d\epsilon}{2\pi i}\int_{-\infty}^{+\infty}\frac{d\epsilon'}{2\pi i}
{\rm Tr}
\left[
{\bm G}(\kv_-, \epsilon){\bm G}(\kv_+, \epsilon+\omega){\bm G}(\kv'_+, \epsilon'+\omega)
{\bm G}(\kv_+, \epsilon +\omega)
\right]\nn\\
& = & -\frac{1}{16}\sum_{\mu,\nu, \lambda,\gamma} \int \frac{d^2 \kv}{(2\pi)^2}\int \frac{d^2 \kv'}{(2\pi)^2} 
V_{\kv-\kv'}n_{\kv'_+,\gamma} \nn\\
&\times&\int_{-\infty}^{+\infty}\frac{d\epsilon}{2\pi i}
\frac{
{\rm Tr}
\left[
(\openone_\sigma + \mu \sigmav_{\kv_-})
(\openone_\sigma + \nu \sigmav_{\kv_+})
(\openone_\sigma + \gamma \sigmav_{\kv'_+})
(\openone_\sigma + \lambda \sigmav_{\kv_+})\right]
}
{
(\epsilon -\xi_{\kv_-, \mu} + i\eta_{\kv_-, \mu})
(\epsilon +\omega - \xi_{\kv_+, \nu} + i\eta_{\kv_+,\nu})
(\epsilon +\omega - \xi_{\kv_+, \lambda} + i\eta_{\kv_+,\lambda})
}~,
\eer
It is easy to show that for $\lambda=\nu$ the trace becomes
\be\label{tr_se-c}
\left. {\rm Tr}
\left[
(\openone_\sigma + \mu \sigmav_{\kv_-})
(\openone_\sigma + \nu \sigmav_{\kv_+})
(\openone_\sigma + \gamma \sigmav_{\kv'_+})
(\openone_\sigma + \lambda \sigmav_{\kv_+})\right]
\right|_{\lambda=\nu} = 16~{\cal F}_{\mu\nu}(\kv_+,\kv_-)~{\cal F}_{\gamma\nu}(\kv_+,\kv'_+)~,
\ee
\end{widetext}
and that the contribution arising from $\lambda=-\nu$ averages to zero after performing the integrations over $\kv$ and $\kv'$. 
Inserting Eq.~(\ref{tr_se-c}) in Eq.~(\ref{chin_se1}) we find that
\begin{widetext}
\ber\label{chin_se1dp}
\chi^{({\rm SE}-{\rm c})}_{\rho\rho}(q,\omega)
& = & -\sum_{\mu, \nu,\gamma} \int \frac{d^2 \kv}{(2\pi)^2}\int \frac{d^2 \kv'}{(2\pi)^2} V_{\kv-\kv'}n_{\kv'_+,\gamma}
\int_{-\infty}^{+\infty}\frac{d\epsilon}{2\pi i}
\frac{{\cal F}_{\mu\nu}(\kv_+, \kv_-){\cal F}_{\gamma\nu}(\kv_+, \kv'_+)}
{(\epsilon - \xi_{\kv_-, \mu} + i \eta_{\kv_-, \mu})
(\epsilon + \omega - \xi_{\kv_+, \nu} + i \eta_{\kv_+,\nu})^2} \nn\\
& = & \partial_{\omega}\sum_{\mu,\nu, \gamma} \int \frac{d^2 \kv}{(2\pi)^2}\int \frac{d^2 \kv'}{(2\pi)^2} 
V_{\kv-\kv'} n_{\kv'_+,\gamma}
{\cal F}_{\mu\nu}(\kv_+,\kv_-){\cal F}_{\gamma\nu}(\kv_+,\kv'_+)
\frac{n_{\kv_-, \mu} - n_{\kv_+, \nu}}{\omega + \xi_{\kv_-, \mu} - \xi_{\kv_+, \nu} + i\eta}~.
\eer
\end{widetext}
For the diagram in Fig.~\ref{fig:two}d) we find that
\begin{widetext}
\ber\label{chin_se2dp}
\chi^{({\rm SE}-{\rm d})}_{\rho\rho}(q,\omega)
& = & -\partial_{\omega}\sum_{\mu, \nu, \gamma} \int \frac{d^2 \kv}{(2\pi)^2}\int \frac{d^2 \kv'}{(2\pi)^2} 
V_{\kv-\kv'} n_{\kv'_-,\gamma}
{\cal F}_{\mu\nu}(\kv_-,\kv_+){\cal F}_{\gamma\nu}(\kv_-,\kv'_-)
\frac{n_{\kv_-, \nu} - n_{\kv_+, \mu}}{\omega + \xi_{\kv_-, \nu} - \xi_{\kv_+, \mu}+i\eta}~.
\eer
\end{widetext}
We now sum Eqs.~(\ref{chin_se1dp}) and (\ref{chin_se2dp}) together and separate inter-band ($\nu=-\mu$) from intra-band ($\nu=\mu$) terms. In the inter-band channel ${\cal F}_{\mu\nu}(\kv_+,\kv_-)$ is already of order $q^2$, and thus the other factors can be calculated at $q=0$:
\begin{widetext}
\be\label{chin_sedpinter}
\chi^{({\rm SE}-{\rm inter})}_{\rho\rho}(q,\omega) =  
- q^2\partial_\omega\sum_{\mu} \int \frac{d^2 \kv}{(2\pi)^2}\int \frac{d^2 \kv'}{(2\pi)^2}
V_{\kv-\kv'}\sin^2(\phi_\kv)\cos(\phi_{\kv'}-\phi_{\kv}) \frac{(1-n_{\kv', +})(1-n_{\kv, +})}{k^2 (\omega + 2\mu v k + i\eta)}~.
\ee
\end{widetext}
The real part of this expression vanishes for $\omega \rightarrow 0$, but its imaginary part gives Eq.~(\ref{ImchiSE}).
On the other hand the intra-band contribution is entirely real and gives Eq.~(\ref{RechiSE}).

\section{Generalized continuity equation for an interacting system of massless Dirac fermions}
\label{app:general-continuity-equation}

In the presence of a rigid momentum cutoff, the continuity equation (\ref{contin}) needs to be modified. To show this we start by introducing field operators ${\h \Psi}_{\kv,\alpha}$ in a restricted Hilbert space (RHS):
\be
{\h \Psi}_{\kv,\alpha}=\Theta(k_{\rm max}- k)\h{\psi}_{\kv,\alpha}~,
\ee
where $k_{\rm max}$ is the ultraviolet momentum cutoff and $\h{\psi}_{\kv,\alpha}$ is the regular field operator.
All other operators should be defined in terms of these new field operators, {\it e.g.} the density operator reads
\be\label{eq:rho_rhs}
{\h \rho}_\qv=\sum_{\kv, \alpha}{\hdg \Psi}_{\kv-\qv, \alpha}{\h \Psi}_{\kv,\alpha}~.
\ee
Now an interesting observation is that the commutator $[{\h \rho}_{\qv}, {\h \rho}_{\qv'}]$, which is zero in the regular space, becomes finite in the RHS:
\ber\label{eq:rhowithrho}
[{\h \rho}_{\qv}, {\h \rho}_{\qv'}] &=& \sum_{\kv, \kv', \alpha,\beta}
[{\hdg \Psi}_{\kv-\qv, \alpha}{\h \Psi}_{\kv,\alpha}, {\hdg \Psi}_{\kv'-\qv',\beta}{\h \Psi}_{\kv',\beta}] \nonumber\\
&=& \sum_{\kv,\alpha}
\left[\Theta(k_{\rm max} - |\kv-\qv'|)
-\{\qv'\to\qv\}\right] \nonumber\\
&\times&{\hdg \Psi}_{\kv-\qv-\qv',\alpha}{\h \Psi}_{\kv, \alpha}~.
\eer
This is not zero in general. One can show that its expectation value $\langle \dots \rangle_0$ over the {\it noninteracting} ground state is zero: 
$\langle[{\h \rho}_{\qv}, {\h \rho}_{\qv'}]\rangle_0 = 0$. It is also straightforward to show that the commutator of the density operator with the kinetic part of the Hamiltonian, ${\hat {\cal H}}_{\rm D}$, remains unchanged
\be
[\h{\rho}_{\qv},{\hat {\cal H}}_{\rm D}]=\qv\cdot\h{\bm j}_{\qv}=v q {\h \sigma}_{\qv}~,
\ee
where the Dirac-Weyl Hamiltonian ${\hat {\cal H}}_{\rm D}$, the current-density operator $\h{\bm j}_{\qv}$, and the longitudinal 
component of the pseudospin-density operator $ \h{\sigma}_{\qv}$ are also redefined in the RHS.

As a consequence of Eq.~(\ref{eq:rhowithrho}), the commutator of ${\h \rho}_{\qv}$ with the Coulomb interaction 
${\hat {\cal H}}_{\rm C}$ is non-zero in the RHS:
\begin{widetext}
\ber\label{fq}
{\h \Gamma}_\qv &=& [{\h \rho}_\qv, {\h {\cal H}}_{\rm C}] = 
\frac{1}{2S}\sum_{\qv'\ne 0}V_{q'}\sum_{\kv, \kv', \pv, \alpha, \beta,\gamma}
[{\hdg \Psi}_{\pv-\qv, \gamma}{\h \Psi}_{\pv, \gamma}, {\hdg \Psi}_{\kv-\qv', \alpha} {\hdg \Psi}_{\kv'+\qv', \beta} 
{\h \Psi}_{\kv', \beta} {\h \Psi}_{\kv, \alpha}] \nonumber \\
& = &\frac{1}{S}\sum_{\qv'\ne0}V_{q'}
\sum_{\kv,\alpha}\sum_{\kv',\beta}
\left[\Theta(k_{\rm max}-|\kv-\qv'|)-\{\qv'\to\qv\}\right]
{\hdg \Psi}_{\kv - \qv - \qv', \alpha}{\hdg \Psi}_{\kv'+\qv', \beta} {\h \Psi}_{\kv', \beta} {\h \Psi}_{\kv, \alpha}~.
\eer
\end{widetext}
Again, one can show that the expectation value of ${\h \Gamma}_\qv$ over the noninteracting 
ground state vanishes: $\langle\h{\Gamma}_\qv\rangle_0=0$.

Now applying twice the following identity~\cite{Giuliani_and_Vignale}
\be
\langle\langle {\h A}; {\h B}\rangle\rangle_\omega=\frac{1}{\omega}\langle[\h{A},\h{B}]\rangle
+\frac{1}{\omega}\langle\langle [\h{A},\h{\cal H}];\h{B}\rangle\rangle_\omega~,
\ee
where
\be
\langle\langle {\h A}; {\h B}\rangle\rangle_\omega = - i \int_0^\infty dt~\langle [{\h A}(t),{\h B}(0)]\rangle e^{-i\omega t} e^{-\eta t}
\ee
is the usual Kubo product~\cite{Giuliani_and_Vignale}, to the density-density response function (${\h A}={\h \rho}_\qv$, ${\h B}={\h \rho}_{-\qv}$), one gets
\ber\label{eom}
\omega^2\langle\langle {\h \rho}_{\qv}; {\h \rho}_{-\qv}\rangle\rangle_{\omega}
&=& \omega\langle[{\h \rho}_{\qv}, {\h \rho}_{-\qv}]\rangle + v q\langle[{\h \sigma}_{\qv}, {\h \rho}_{-\qv}]\rangle \nonumber\\
&+& \langle[{\h \Gamma}_\qv, {\h \rho}_{-\qv}]\rangle + v^2 q^2 \langle\langle {\h \sigma}_{\qv}; {\h \sigma}_{-\qv}\rangle\rangle_{\omega}\nonumber\\
&+& v q \langle\langle {\h \sigma}_\qv; {\h \Gamma}_{-\qv}\rangle\rangle_{\omega}
 - v q \langle\langle {\h \Gamma}_{\qv}; {\h \sigma}_{-\qv}\rangle\rangle_{\omega}\nonumber\\
& - & \langle\langle {\h \Gamma}_{\qv}; {\h \Gamma}_{-\qv}\rangle\rangle_{\omega}~.
\eer
We remind the reader that the linear-response function $\chi_{AB}(\omega)$ is directly related to the Kubo product by the relation 
$\chi_{AB}(\omega) =\langle\langle {\h A}; {\h B}\rangle\rangle_\omega / S$. In the noninteracting limit the first term on the r.h.s. of Eq.~(\ref{eom}) is identically zero, the second term is the anomalous commutator~\cite{sabio_prb_2008}, and all terms involving ${\h \Gamma}_\qv$ vanish.

Here, we are mainly interested in the imaginary part of Eq.~(\ref{eom}):
\ber\label{im_eom}
\Im m~\chi_{\rho\rho}(q, \omega) &= &\frac{v^2q^2}{\omega^2}\Im m~\chi_{\sigma\sigma}(q, \omega) \nonumber \\
&+& \frac{vq}{\omega^2}\left[\Im m~\chi_{\sigma\Gamma}(q, \omega) - \Im m~\chi_{\Gamma \sigma}(q, \omega)\right] \nonumber \\
&+& \frac{1}{\omega^2}\Im m~\chi_{\Gamma \Gamma}(q, \omega)~.
\eer
The last term on the r.h.s. of Eq.~(\ref{im_eom}) is at least of second order in the electron-electron interaction. Thus, up to first order in the Coulomb interaction, we can write
\ber\label{im_eom1}
\Im m~\chi^{(1)}_{\rho\rho}(q, \omega) &=& \frac{v^2 q^2}{\omega^2}\Im m~\chi^{(1)}_{\sigma\sigma}(q,\omega) \nonumber\\
&+& \frac{vq}{\omega^2}\left[\Im m~\chi^{(0)}_{\sigma\Gamma}(q, \omega) - \Im m~\chi^{(0)}_{\Gamma \sigma}(q, \omega)\right]~,\nonumber\\
\eer
where the superscript ``$(n)$" on the response functions in the previous equation indicates that they have to be evaluated up to the
$n$-th order in the electron-electron interaction.

In order to evaluate $\chi^{(0)}_{\sigma\Gamma}(q, \omega)$ and $\chi^{(0)}_{\Gamma\sigma}(q, \omega)$ we start from the well known exact-eigestate representation expression for a response function at zero temperature~\cite{Giuliani_and_Vignale}:
\be
\chi_{AB}(\omega)=\sum_n\left[
\frac{\bra{0} \h{A}\ket{n}\bra{n}\h{B}\ket{0}}{\omega-\omega_{n0}+i\eta}
-\frac{\bra{0} \h{B}\ket{n}\bra{n}\h{A}\ket{0}}{\omega+\omega_{n0}+i\eta}
\right]~.
\ee
Before proceeding further let us also introduce the following unitary transformation which diagonalizes the Dirac-Weyl Hamiltonian 
${\h {\cal H}}_{\rm D}$:
\be
\left\{ 
\begin{array}{l}
{\h \Psi}_{\kv, \alpha} = \sum_\mu {\cal U}_{\alpha\mu}(\kv){\h c}_{\kv, \mu}\vspace{0.1 cm}\\
{\hdg \Psi}_{\kv, \alpha} = \sum_\mu {\cal U}^\dagger_{\mu\alpha}(\kv) {\hdg c}_{\kv, \mu}
\end{array} 
\right.~,
\ee
where
\be
{\cal U}(\kv)=\frac{1}{\sqrt{2}}
\begin{pmatrix} 
e^{-i\phi_\kv/2}  & e^{-i\phi_\kv/2} \\
e^{i\phi_\kv/2} & -e^{i\phi_\kv/2} 
\end{pmatrix}~.
\ee
The full Hamiltonian ${\cal H}$ after this unitary transformation reads
\begin{widetext}
\be\label{hmlt_bnd}
{\h {\cal H}}'= \sum_{\kv,\mu}\varepsilon_{\kv, \mu} {\hdg c}_{\kv, \mu}{\h c}_{\kv, \mu}
+\frac{1}{2S}\sum_{\qv \neq {\bm 0}} V_{q}\sum_{\kv, \kv', \mu,\nu, \mu',\nu'}
I^{\mu\nu}_{\kv, \kv+\qv} I^{\mu'\nu'}_{\kv', \kv'-\qv}
{\hdg c}_{\kv, \mu}{\hdg c}_{\kv', \mu'} {\h c}_{\kv'-\qv,\nu'}{\h c}_{\kv+\qv,\nu}~,
\ee
\end{widetext}
where $\varepsilon_{\kv, \pm} = \pm vk$ are Dirac-band energies and the matrix elements $I^{\mu\nu}_{\kv, \kv'}$ are defined as
\ber\label{f}
I^{\mu\nu}_{\kv,\kv'} &=& \left[{\cal U}^{\dagger}(\kv){\cal U}(\kv')\right]_{\mu\nu}\\
&=& \frac{e^{i (\phi_{\kv} - \phi_{\kv'}) / 2}+ \mu\nu e^{-i (\phi_{\kv} - \phi_{\kv'})/2}}{2}~.
\eer
We also introduce
\ber\label{g}
X^{\mu\nu}_{\kv,\kv'} &=& \left[{\cal U}^{\dagger}(\kv)\sigma^x {\cal U}(\kv')\right]_{\mu\nu} \nonumber\\
& = & \frac{\mu e^{-i (\phi_{\kv}+\phi_{\kv'})/2} + \nu e^{i\left(\phi_{\kv}+\phi_{\kv'}\right)/2}}{2}~.
\eer
Now we can write
\begin{widetext}
\ber
\chi^{(0)}_{\sigma\Gamma}(q,\omega) &=&
g\sum_n\left[
\frac{\bra{0} \h{\sigma}^x_{\qv}\ket{n}_0 \bra{n}\h{\Gamma}_{-\qv}\ket{0}_0}{\omega-\omega_{n0}+i\eta}
-\frac{\bra{0}\h{\Gamma}_{-\qv}\ket{n}_0 \bra{n}\h{\sigma}^x_{\qv}\ket{0}_0}{\omega+\omega_{n0}+i\eta}
\right] \nonumber\\
&=& g\sum_{\pv, \gamma, \lambda} X^{\gamma\lambda}_{\pv - \qv, \pv}
\frac{\langle {\hdg c}_{\pv-\qv, \gamma}{\h c}_{\pv, \lambda}{\h \Gamma}_{-\qv}\rangle_0 - \langle {\h \Gamma}_{-\qv}
{\hdg c}_{\pv-\qv,\gamma}{\h c}_{\pv, \lambda}\rangle_0}{\omega + \varepsilon_{\pv-\qv, \gamma} - \varepsilon_{\pv, \lambda} + i\eta}~.
\eer
\end{widetext}
Using the expression of ${\h \Gamma}_\qv$ given above in Eq.~(\ref{fq}) and Wick's theorem~\cite{Giuliani_and_Vignale}, and keeping only terms linear in $q$ we finally find that
\begin{widetext}
\be\label{sigmaGamma_q}
\chi^{(0)}_{\sigma\Gamma}(q, \omega) = \frac{qg}{S^2}\sum_{\lambda,\nu}\sum_{\kv,\pv < k_{\rm max}} 
X^{{\bar \lambda} \lambda}_{\pv, \pv} I^{\lambda \nu}_{\pv, \kv} I^{\nu {\bar \lambda}}_{\kv, \pv}
V_{\pv-\kv}n_{\kv, \nu}
\left[2\cos(\phi_\kv)\delta(k_{\rm max}-k|) + \cos(\phi_\pv)\delta(k_{\rm max}-p)\right]
\frac{n_{\pv, \lambda} - n_{\pv, {\bar \lambda}}}{\omega - 2\varepsilon_{\pv, \lambda} + i\eta}~.
\ee
\end{widetext}
In deriving this expression we have assumed that the Coulomb interaction is screened ({\it i.e.} that $V_q$ is not singular for $q \to 0$).

The imaginary part of Eq.~(\ref{sigmaGamma_q}) for $0< \omega< 2v k_{\rm max}$ reads
\begin{widetext}
\ber
\Im m~\chi^{(0)}_{\sigma\Gamma}(q,\omega) &=& 
\frac{g\pi q}{S^2} \sum_{\kv, \pv < k_{\rm max}} V_{\pv-\kv}
(1-n_{\kv, +})(1-n_{\pv, +})
\sin(\phi_\pv)\sin(\phi_\pv-\phi_\kv)\cos(\phi_\kv) \delta(k_{\rm max} - k) 
\delta(\omega-2 v p)\nonumber \\
& = & \frac{q k_{\rm max} \omega}{32\pi v^2}\Theta(\omega-2\varepsilon_{\rm F})
[V_0(\omega/(2 v), k_{\rm max}) -  V_2(\omega/(2 v), k_{\rm max})]~.
\eer
\end{widetext}
Using a similar expression for $\Im m~\chi^{(0)}_{\Gamma\sigma}(q,\omega)$ we finally obtain
\begin{widetext}
\be
\Im m~\chi^{(0)}_{\sigma\Gamma}(q,\omega) - \Im m~\chi^{(0)}_{\Gamma\sigma}(q,\omega) = 
\frac{q k_{\rm max} \omega}{2^4 \pi v^2}\Theta(\omega - 2\varepsilon_{\rm F})
[V_0(\omega/(2 v), k_{\rm max}) - V_2(\omega/(2 v), k_{\rm max})]~.
\ee
\end{widetext}
Taking the limit $k_{\rm max}\to \infty$ the previous equation simplifies considerably to
\be
\Im m~\chi^{(0)}_{\sigma\Gamma}(q, \omega) - \Im m~\chi^{(0)}_{\Gamma\sigma}(q, \omega) 
= \frac{q \omega \alpha_{\rm ee}}{8 v}\Theta(\omega - 2\varepsilon_{\rm F})~.
\ee
Using this result in Eq.~(\ref{im_eom1}) we find
\be
\Im m~\chi^{(1)}_{\rho\rho}(q, \omega) = \frac{v^2 q^2}{\omega^2}\Im m~\chi^{(1)}_{\sigma\sigma}(q,\omega)
+ \frac{\alpha_{\rm ee}q^2}{8 \omega}\Theta(\omega - 2\varepsilon_{\rm F})~,
\ee
or, in terms of the a.c. conductivity,
\be\label{sigma1_xFFx}
\Re e~\sigma^{(1)}(\omega) = -\frac{e^2 v^2}{\omega}\Im m~\chi^{(1)}_{\sigma\sigma}(\omega) - \frac{\alpha_{\rm ee}}{2}\sigma_0~.
\ee
\section{Numerical results for Thomas-Fermi screened interactions}
\label{app:thomasfermi}

\begin{figure}
\includegraphics[width=1.0\linewidth]{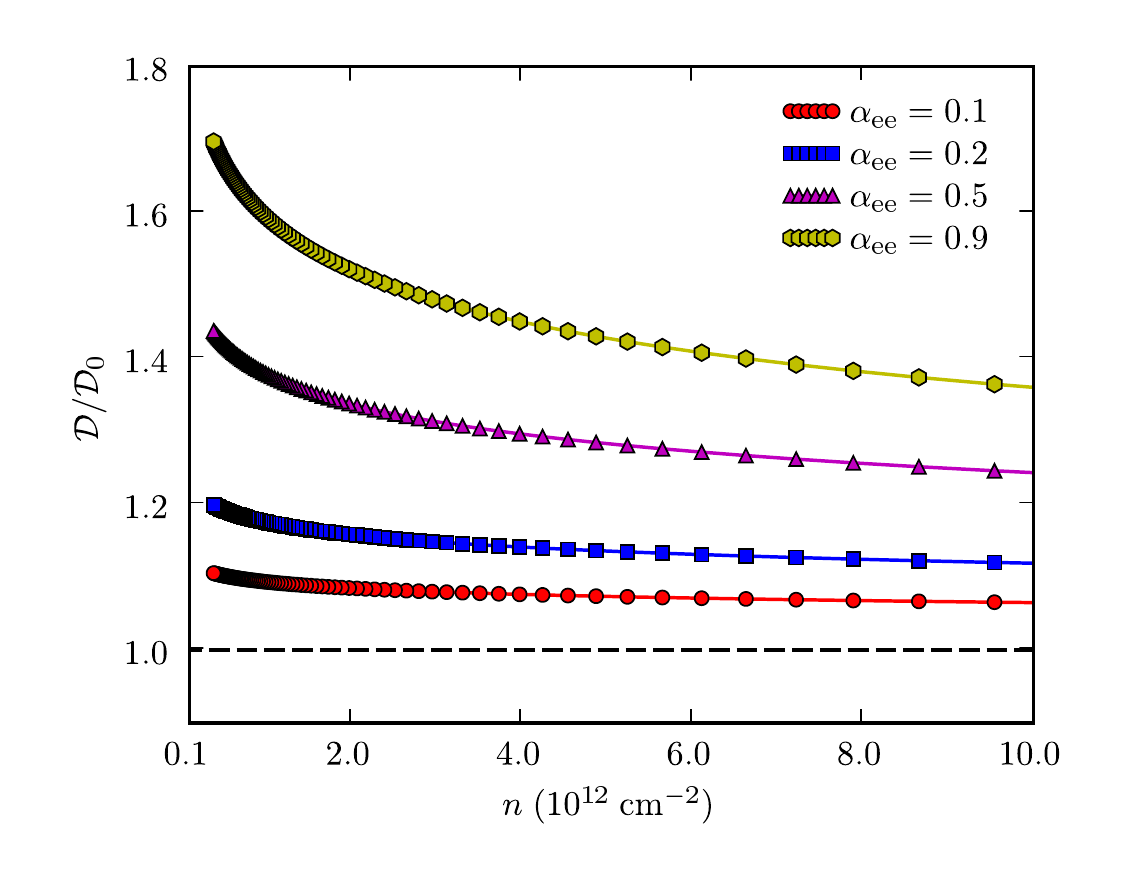} 
\caption{(Color online) Same as in Fig.~\ref{fig:three} but for Thomas-Fermi screened interactions.\label{fig:six}}
\end{figure}

\begin{figure}
\includegraphics[width=1.0\linewidth]{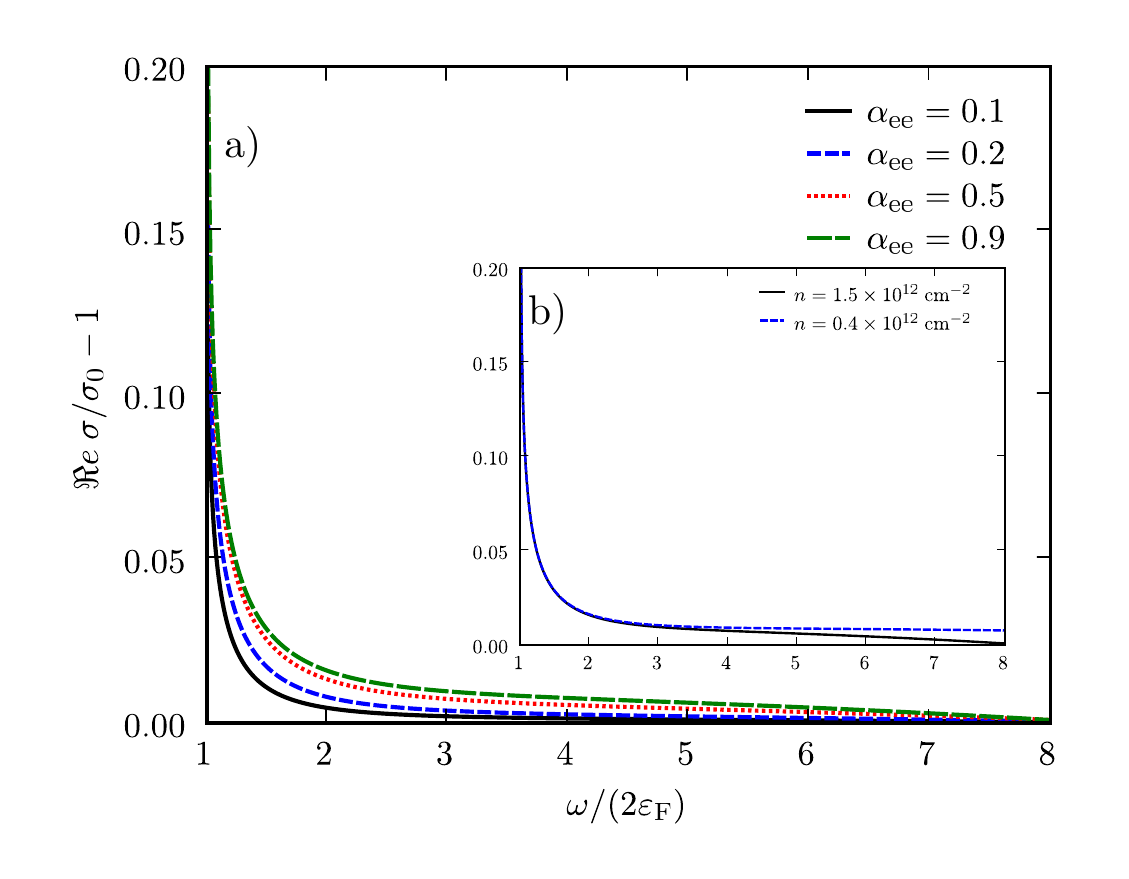} 
\caption{(Color online) Same as in Fig.~\ref{fig:four} but for Thomas-Fermi screened interactions. \label{fig:seven}}
\end{figure}

\begin{figure}
\includegraphics[width=1.0\linewidth]{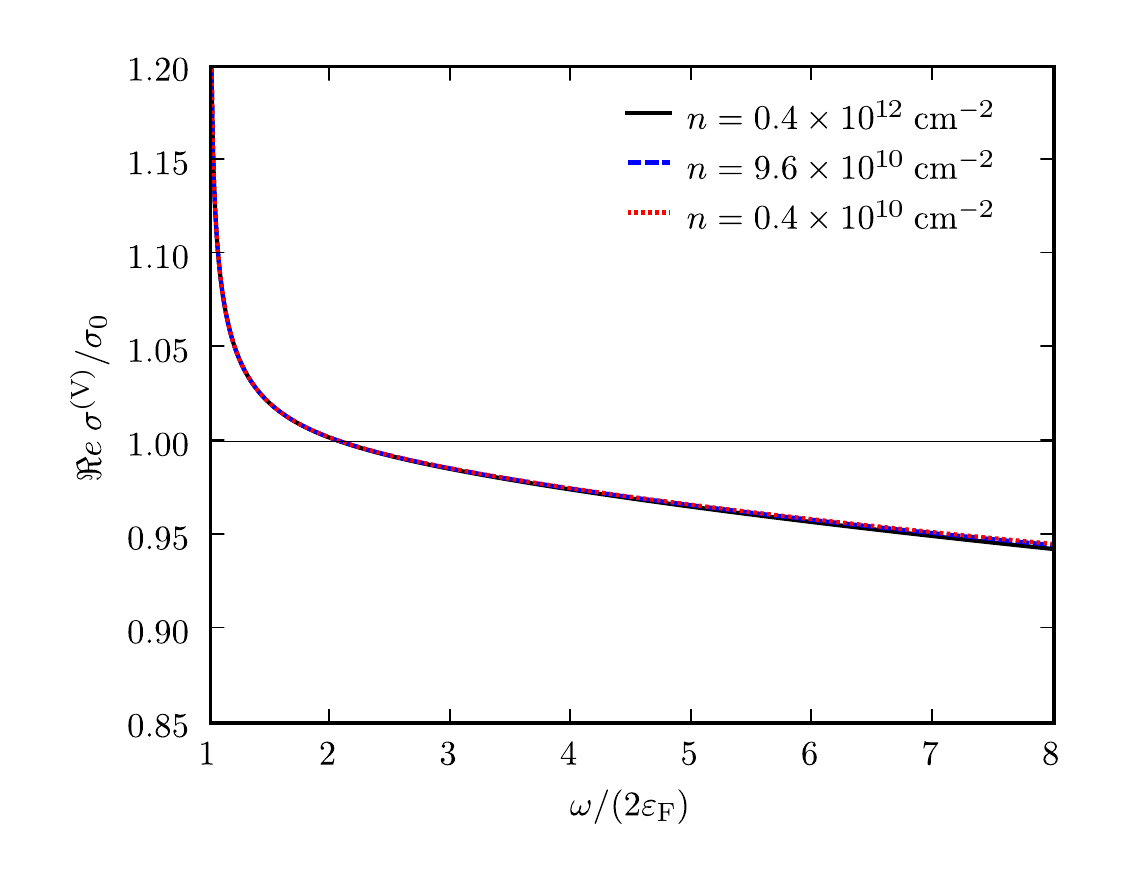} 
\caption{(Color online) Same as in Fig.~\ref{fig:five} but for Thomas-Fermi screened interactions. Note that the scale in the vertical axis in this figure is completely different from that in Fig.~\ref{fig:five}.\label{fig:eight}}
\end{figure}

In Figs.~\ref{fig:six}-\ref{fig:eight} we present numerical results for ${\cal D}$, $\Re e~\sigma(\omega)$, and $\Re e~\sigma^{({\rm V})}(\omega)$ obtained by using 
Thomas-Fermi screened interactions, {\it i.e.} $q_{\rm TF} \neq 0$ in Eq.~(\ref{eq:Coulomb_pseudopotentials}).



\begin{thebibliography}{77}

\bibitem{tonks_langmuir_pr_1929}
	L. Tonks  and I. Langmuir, {Phys. Rev.} {\bf 33}, 195 (1929).
	
\bibitem{pines_bohm_pr_1952}
	D. Pines and D. Bohm, {Phys. Rev.} {\bf 85}, 338 (1952).

\bibitem{Pines_and_Nozieres}
	D. Pines and P. Nozi\'eres,  {\it The Theory of Quantum Liquids} (W.A. Benjamin, Inc., New York, 1966).

\bibitem{Ebbesen_PT_2008}
	 T.W. Ebbesen, C. Genet,  and S.I. Bozhevolnyi, { Phys. Today} {\bf  61}(5), 44 (2008).
	
\bibitem{Maier07} 
	 S.A. Maier,  {\it Plasmonics -- Fundamentals and Applications} (Springer, New York, 2007). 

\bibitem{Giuliani_and_Vignale}
	G.F. Giuliani and G. Vignale, {\it Quantum Theory of the Electron Liquid} (Cambridge University Press, Cambridge, 2005).	

\bibitem{morchio_strocchi_ap_1986}
	G. Morchio and F. Strocchi, Ann. Phys. (NY) {\bf 170}, 310 (1986).
	
\bibitem{Cardona}  
	P.Y. Yu and M. Cardona, {\it Fundamentals of Semiconductors} (Springer-Verlag, Berlin, 1999).
	
\bibitem{vittorio_ils}
	For a recent review see {\it e.g.} V. Pellegrini and A. Pinczuk,  Phys. Stat. Sol. (B) {\bf 243}, 3617 (2006).

\bibitem{Hirjibehedin_prb_2007}
	C.F. Hirjibehedin, A. Pinczuk, B.S. Dennis, L.N. Pfeiffer, and K.W. West, \prb {\bf 65}, 161309 (2002).

\bibitem{geim_novoselov_nat_mat_2007} 
	A.K. Geim,  and K.S. Novoselov, Nature Mater. {\bf 6}, 183 (2007).

\bibitem{katsnelson_ssc_2007}
	M.I. Katsnelson and  K.S. Novoselov, Solid. State Commun. {\bf 143}, 3 (2007).

\bibitem{allan_pt_2007} 
	A.K. Geim and A.H. MacDonald, Phys. Today {\bf 60}, 35 (2007).

\bibitem{castro_neto_rmp_2009}
	A.H. Castro Neto, F. Guinea, N.M.R. Peres, K.S. Novoselov, and A.K. Geim,  \rmp {\bf 81}, 109 (2009).

\bibitem{kotov_arXiv_2010}
	For recent reviews on electron-electron interactions in graphene see {\it e.g.} 
	D.S.L. Abergel, V. Apalkov, J. Berashevich, K. Ziegler, and T. Chakraborty, Adv. Phys. {\bf 59}(4), 261 (2010) and 
	V.N. Kotov, B. Uchoa, V.M. Pereira, A.H. Castro Neto, and F. Guinea, arXiv:1012.3484.

\bibitem{electron_doping}
	We discuss electron doping for the sake of definiteness.  The graphene properties discussed in this paper are particle-hole symmetric.	

\bibitem{Tse_KE} 
	W.-K. Tse {\it et al.}, to be submitted for publication.   
	
\bibitem{proper_diagrams}
	${\widetilde \chi}_{\rho\rho}(q,\omega) = \chi_{\rho\rho}(q,\omega)\varepsilon(q,\omega)$, 
	where $\chi_{\rho\rho}(q,\omega)$ is the usual physical (causal) density-density response function 
	and $\varepsilon(q,\omega)$ is the dielectric function. Physically ${\widetilde \chi}_{\rho\rho}(q,\omega)$ describes 
	the response to the screened potential and is defined diagramatically 
	as the sum of all the ``proper" diagrams$^{\,}$\cite{Giuliani_and_Vignale}, 
	{\it i.e.} those diagrams that cannot be separated into two parts, each one containing one of the external vertices, 
	by the cutting of a single interaction line.
	
\bibitem{kuzmenko_prl_2008}
	A.B. Kuzmenko, E. van Heumen, F. Carbone, and D. van der Marel, \prl {\bf 100}, 117401 (2008).

\bibitem{nair_science_2008} 
	R.R. Nair, P. Blake, A.N. Grigorenko, K.S. Novoselov, T.J. Booth, T. Stauber, N.M.R. Peres,  and A.K. Geim,  
	Science {\bf 320}, 1308 (2008).

\bibitem{wang_science_2008}
	F. Wang, Y. Zhang, C. Tian, C. Girit, A. Zettl, M. Crommie, and Y.R. Shen, Science {\bf 320}, 206 (2008).
	
\bibitem{li_natphys_2008}
	Z.Q. Li,  E.A. Henriksen, Z. Jiang, Z. Hao, M.C. Martin, P. Kim, H.L. Stormer, and D.N. Basov, Nature Phys. {\bf 4}, 532 (2008).

\bibitem{mak_prl_2008}
	K.F. Mak, M.Y. Sfeir, Y. Wu, C.H. Lui,  J.A.  Misewich, and  T.F. Heinz, Phys. Rev. Lett. {\bf 101}, 196405 (2008).

\bibitem{jang_prl_2008}
	C. Jang,  S. Adam, J.-H. Chen, E.D. Williams, S. Das Sarma, and M.S. Fuhrer, {Phys. Rev. Lett.} {\bf 101}, 146805 (2008).

\bibitem{mohiuddin_preprint_2008}
	L.A. Ponomarenko, R. Yang, T.M. Mohiuddin, M.I. Katsnelson, K.S. Novoselov, S.V. Morozov, 
	A.A. Zhukov, F. Schedin, E.W. Hill, and A.K. Geim, Phys. Rev. Lett. {\bf 102}, 206603 (2009).

\bibitem{wunsch_njp_2006}
	B. Wunsch, T. Stauber, F. Sols, and  F.  Guinea, { New J. Phys.} {\bf 8}, 318 (2006).

\bibitem{hwang_prb_2007}
	E.H. Hwang, and S. Das Sarma, {Phys. Rev. B} {\bf 75}, 205418 (2007).

\bibitem{polini_prb_2008}
	M. Polini, R. Asgari, G. Borghi, Y. Barlas, T. Pereg-Barnea, and A.H. MacDonald, \prb {\bf 77}, 081411(R) (2008).
	
\bibitem{principi_prb_2009}
	A. Principi, M. Polini, and G. Vignale, \prb {\bf 80}, 075418 (2009).

\bibitem{yacoby_natphys_2008} 
	J. Martin, N. Akerman, G. Ulbricht, T. Lohmann, J.H. Smet, K. von Klitzing, and A. Yacoby, Nature Phys. {\bf 4}, 144 (2008).

\bibitem{mishchenko_prl_2007}
	E.G. Mishchenko, \prl {\bf 98}, 216801 (2007).
	
\bibitem{borghi_ssc_2009}
	G. Borghi, M. Polini, R. Asgari, and A.H. MacDonald, Solid State Commun. {\bf 149}, 1117 (2009).	

\bibitem{gonzalez_nuclearphys_1994} 
	J. Gonz\'alez, F. Guinea, and M.A.H. Vozmediano, Nucl. Phys. B {\bf 424}, 595 (1994); 
	\prb {\bf 59}, R2474 (1999).

\bibitem{barlas_prl_2007}
	Y. Barlas, T. Pereg-Barnea, M. Polini, R. Asgari, and A.H. MacDonald, \prl {\bf 98}, 236601 (2007);
	M. Polini, R. Asgari, Y. Barlas, T. Pereg-Barnea, and A.H. MacDonald, Solid State Commun. {\bf 143}, 58 (2007).
	
\bibitem{Tse_PRB_2007} 
	S. Das Sarma, E.H. Hwang, and W.-K. Tse, Phys. Rev. B {\bf 75}, 121406(R) (2007).

\bibitem{polini_unpublished}
	M. Polini, A.H. MacDonald, and G. Vignale, unpublished.

\bibitem{sabio_prb_2008}
	J. Sabio, J. Nilsson, and A.H. Castro Neto, \prb {\bf 78}, 075410 (2008).
	
\bibitem{giamarchi_book}
	See for example T. Giamarchi, {\it Quantum Physics in One Dimension} (Clarendon Press, Oxford, 2004).

\bibitem{mishchenko_epl_2008}
	E.G. Mishchenko, Europhys. Lett. {\bf 83}, 17005 (2008).
	
\bibitem{sheehy_prl_2007} 
	D.E. Sheehy and J. Schmalian, \prl {\bf 99}, 226803 (2007).
	
\bibitem{herbut_prl_2008}
	I.F. Herbut, V. Juri\v ci\' c, and O. Vafek, \prl {\bf 100}, 046403 (2008).

\bibitem{katsnelson_epl_2008}
	M.I. Katsnelson, Europhys. Lett. {\bf 84}, 37001 (2008).
	
\bibitem{sheehy_prb_2009}
	D.E. Sheehy and J. Schmalian, \prb {\bf 80}, 193411 (2009).
	
\bibitem{polini_arxiv_2009}
	M. Polini, A.H. MacDonald, and G. Vignale, arXiv:0901.4528 (unpublished).

\bibitem{Berkeley}
	J. Horng, C.-F. Chen, B. Geng, C. Girit, Y. Zhang, Z. Hao, H.A. Bechtel, M. Martin, A. Zettl, M.F. Crommie, Y.R. Shen, 
	and F. Wang, arXiv:1007.4623.
	
\bibitem{peres_prl_2010}
	N.M.R. Peres, R.M. Ribeiro, and A.H. Castro Neto, \prl {\bf 105}, 055501 (2010).
	Excitonic effects have also been studied by {\it ab~initio} methods in 
	Li Yang, J. Deslippe, C.-H. Park, M.L. Cohen, and S.G. Louie, \prl {\bf 103}, 186802 (2009).
	Electron-electron interaction effects within a phenomenological self-energy model based on the ``marginal Fermi liquid" 
	concept have been studied in A.G. Grushin, B. Valenzuela, and M.A.H. Vozmediano, \prb {\bf 80}, 155417 (2009).

\bibitem{theoryworkopticalconductivity}
	N.M.R. Peres, F. Guinea, and A.H. Castro Neto, \prb {\bf 73}, 125411 (2006);
	L.A. Falkovsky and S.S. Pershoguba, {\it ibid.} {\bf 76}, 153410 (2007);
	T. Stauber, N.M.R. Peres, and A.H. Castro Neto, {\it ibid.} {\bf 78}, 085418 (2008);
	T. Stauber, N.M.R. Peres, and A.K. Geim, {\it ibid.} {\bf 78}, 085432 (2008);
	V.P. Gusynin, S.G. Sharapov, and  J.P. Carbotte, New J. Phys. {\bf 11}, 095013 (2009);
	J.P. Carbotte, E.J. Nicol, and S.G. Sharapov, \prb {\bf 81}, 045419 (2010);
	K. Jahanbani and R. Asgari, Eur. Phys. J. B {\bf 73}, 247 (2010);
	F.M.D. Pellegrino, G.G.N. Angilella, and R. Pucci, \prb {\bf 81}, 035411 (2010).

\bibitem{kadanoff-baym}
	L.P. Kadanoff and G. Baym, {\it Quantum Statistical Mechanics} (W.A. Benjamin, New York, 1962).
	
\bibitem{tse_prb_2009}
	W.-K. Tse and A.H. MacDonald, \prb {\bf 80}, 195418 (2009).	
	
\bibitem{jiang_prl_2007}
	Z. Jiang, E.A. Henriksen, L.C. Tung, Y.-J. Wang, M.E. Schwartz, M.Y. Han, P. Kim, 
	and H.L. Stormer,  \prl {\bf 98}, 197403 (2007).
	
\bibitem{deacon_prb_2007}
	R.S. Deacon, K.-C. Chuang, R.J. Nicholas, K.S. Novoselov, and A.K. Geim, 
	\prb {\bf 76}, 081406(R) (2007).

\bibitem{henriksen_prl_2010}
	E.A. Henriksen, P. Cadden-Zimansky, Z. Jiang, Z.Q. Li, L.-C. Tung, M.E. Schwartz, M. Takita, Y.-J. Wang, P. Kim, and H.L. Stormer,
	\prl {\bf 104}, 067404 (2010).

\bibitem{iyengar_prb_2007}
	A. Iyengar, J. Wang, H.A. Fertig, and L. Brey, \prb {\bf 75}, 125430 (2007).

\bibitem{bychkov_prb_2008} 
	Yu. A. Bychkov and G. Martinez, \prb {\bf 77}, 125417 (2008).

\bibitem{roldan_prb_2009}
	R. Rold\'an, J.-N. Fuchs, and M.O. Goerbig, \prb {\bf 80}, 085408 (2009).

\bibitem{kohn_pr_1961}
	W. Kohn, Phys. Rev. {\bf 123}, 1242 (1961).

\bibitem{mueller_prb_2008}
	M. M\"uller and S. Sachdev, \prb {\bf 78}, 115419 (2008).
	
\bibitem{agarwal_arXiv_2010}
	A. Agarwal, S. Chesi, T. Jungwirth, J. Sinova, G. Vignale, and M. Polini, arXiv:1010.5169.

\end{thebibliography}
\end{document}